\documentclass[12pt]{iopart}
\usepackage{latexsym}
\usepackage{amscd}
\usepackage[dvips]{color}

\begin{document}
\definecolor{r}{rgb}{1,0,0}
\definecolor{b}{rgb}{0,0,1}
\definecolor{g}{cmyk}{0,0,1,0}
\jl{1}

\jl{1}
 \def\lambdabar{\protect\@lambdabar}
\def\@lambdabar{%
\relax
\bgroup
\def\@tempa{\hbox{\raise.73\ht0
\hbox to0pt{\kern.25\wd0\vrule width.5\wd0
height.1pt depth.1pt\hss}\box0}}%
\mathchoice{\setbox0\hbox{$\displaystyle\lambda$}\@tempa}%
{\setbox0\hbox{$\textstyle\lambda$}\@tempa}%
{\setbox0\hbox{$\scriptstyle\lambda$}\@tempa}%
{\setbox0\hbox{$\scriptscriptstyle\lambda$}\@tempa}%
\egroup
}

\def\bbox#1{%
\relax\ifmmode
\mathchoice
{{\hbox{\boldmath$\displaystyle#1$}}}%
{{\hbox{\boldmath$\textstyle#1$}}}%
{{\hbox{\boldmath$\scriptstyle#1$}}}%
{{\hbox{\boldmath$\scriptscriptstyle#1$}}}%
\else
\mbox{#1}%
\fi
}
\newcommand{\muv}{\bbox{\mu}}
\newcommand{\mc}{{\mathcal M}}
\newcommand{\pc}{{\mathcal P}}
\newcommand{\mct}{\bbox{\mathcal M}}
\newcommand{\pct}{\bbox {\mathcal P}}
\newcommand{\fsf}{{\sf F}}
\newcommand{\fsft}{\bbox{{\sf F}}}
 \newcommand{\mv}{\bbox{m}}
\newcommand{\pv}{\bbox{p}}
\newcommand{\tv}{\bbox{t}}
\def\msf{\hbox{{\sf M}}}
\def\msft{\bbox{{\sf M}}}
\def\psf{\hbox{{\sf P}}}
\def\psft{\bbox{{\sf P}}}
\def\Nsf{\hbox{{\sf N}}}
\def\Nsft{\bbox{{\sf N}}}
\def\Tsf{\hbox{{\sf T}}}
\def\Tsft{\bbox{{\sf T}}}
\def\Asf{\hbox{{\sf A}}}
\def\Asft{\bbox{{\sf A}}}
\def\Bsf{\hbox{{\sf B}}}
\def\Bsft{\bbox{{\sf B}}}
\def\Lsf{\hbox{{\sf L}}}
\def\Lsft{\bbox{{\sf L}}}
\def\Ssf{\hbox{{\sf S}}}
\def\Ssft{\bbox{{\sf S}}}
\def\Mtens{\bi{M}}
\def\msfsim{\bbox{{\sf M}}_{\scriptstyle\rm(sym)}}
\newcommand{\mcsim}{ {\sf M}_{ {\scriptstyle \rm {(sym)} } i_1\dots i_n}}
\newcommand{\mcs}{ {\sf M}_{ {\scriptstyle \rm {(sym)} } i_1i_2i_3}}

\newcommand{\beqan}{\begin{eqnarray*}}
\newcommand{\eeqan}{\end{eqnarray*}}
\newcommand{\beqa}{\begin{eqnarray}}
\newcommand{\eeqa}{\end{eqnarray}}

 \newcommand{\suml}{\sum\limits}
 \newcommand{\sumd}{\suml_{\mathcal D}}
\newcommand{\intl}{\int\limits}
\newcommand{\rvec}{\bbox{r}}
\newcommand{\xivec}{\bbox{\xi}}
\newcommand{\Avec}{\bbox{A}}
\newcommand{\Rvec}{\bbox{R}}
\newcommand{\Evec}{\bbox{E}}
\newcommand{\Bvec}{\bbox{B}}
\newcommand{\Svec}{\bbox{S}}
\newcommand{\avec}{\bbox{a}}
\newcommand{\nablav}{\bbox{\nabla}}
\newcommand{\nuvec}{\bbox{\nu}}
\newcommand{\bvec}{\bbox{\beta}}
\newcommand{\vvec}{\bbox{v}}
\newcommand{\jvec}{\bbox{J}}
\newcommand{\nvec}{\bbox{n}}
\newcommand{\pvec}{\bbox{p}}
\newcommand{\mvec}{\bbox{m}}
\newcommand{\evec}{\bbox{e}}
\newcommand{\eps}{\varepsilon}
\newcommand{\la}{\lambda}
\newcommand{\rad}{\mbox{\footnotesize rad}}
\newcommand{\scr}{\scriptstyle}
\newcommand{\latens}{\bbox{\sf{\Lambda}}}
\newcommand{\lasf}{{\sf{\Lambda}}}
\newcommand{\pitens}{\sf{\Pi}}
\newcommand{\cm}{{\cal M}}
\newcommand{\cp}{{\cal P}}
\newcommand{\beq}{\begin{equation}} 
\newcommand{\eeq}{\end{equation}}
\newcommand{\ptens}{\bbox{\sf{P}}}
\newcommand{\Ptens}{\bbox{P}}
\newcommand{\Ttens}{\bbox{\sf{T}}}
\newcommand{\Ntens}{\bbox{\sf{N}}}
\newcommand{\Ncal}{\bbox{{\cal N}}}
\newcommand{\Atens}{\bbox{\sf{A}}}
\newcommand{\Btens}{\bbox{\sf{B}}}
\newcommand{\dom}{\mathcal{D}}
\newcommand{\al}{\alpha}
\newcommand{\sym}{\scriptstyle \rm{(sym)}}
\newcommand{\Tcal}{\bbox{{\mathcal T}}}
\newcommand{\Nmc}{{\mathcal N}}
\renewcommand{\d}{\partial}
\def\rmi{{\rm i}}
\def\rme{\hbox{\rm e}}
\def\rmd{\hbox{\rm d}}
\newcommand{\ct}{\mbox{\Huge{.}}}
\newcommand{\Laop}{\bbox{\Lambda}}
\newcommand{\Ssfs}{{\scriptstyle \Ssft^{(n)}}}
\newcommand{\Lsfs}{{\scriptstyle \Lsft^{(n)}}}
\newcommand{\psfr}{\widetilde{\psf}}
\newcommand{\msfr}{\widetilde{\msf}}
\newcommand{\msftr}{\widetilde{\msft}}
\newcommand{\psftr}{\widetilde{\psft}}
\newcommand{\pvr}{\widetilde{\pvec}}
\newcommand{\mvr}{\widetilde{\mvec}}
\newcommand{\qdot}{\stackrel{\cdot\cdot\cdot\cdot}}
\newcommand{\bsy}{\hbox}
\newcommand{\ointl}{\oint\limits}
\newcommand{\pisf}{{\sf \Pi}}
\def\Nsf{{\sf N}}
\def\Nsft{\boldsymbol{{\sf N}}}
\newcommand{\gamsf}{{\sf \Gamma}}
\newcommand{\gamsft}{\bsy{\sf \Gamma}}
\newcommand{\ab}{\v{a}} 
\newcommand{\ai}{\^{a}} 
\newcommand{\ib}{\^{\i}} 
\newcommand{\tb}{\c{t}} 
\newcommand{\st}{\c{s}}
\newcommand{\Ab}{\v{A}} 
\newcommand{\Ai}{\^{A}} 
\newcommand{\Ib}{\^{I}} 
\newcommand{\Tb}{\c{T}}
\newcommand{\St}{\c{S}}

\title{Singular behaviour of the electromagnetic field }
\author{C.\ Vrejoiu\footnote{E-mail :  vrejoiu@fizica.unibuc.ro} and R.\ Zus \footnote{E-mail: roxana.zus@fizica.unibuc.ro}}
\address{University of Bucharest, Department of Physics,  \\
PO Box MG - 11, Bucharest-Magurele, RO - 077125,
 Romania }

\begin{abstract}
The singularities of the electromagnetic field  are derived to include all the point-like multipoles representing an electric charge and current distribution. Firstly derived in the static case, the result is generalized to the dynamic one. We establish a simple procedure for passing from the first, to the second case.
\end{abstract}
\section{Introduction}\label{intro}
In the cases of electrostatic and magnetostatic fields of point-like dipoles, one has the well-known procedure of introducing  Dirac $\delta$-function terms for obtaining correct expressions of the electric and magnetic fields defined on the entire space.  The corresponding field expressions take the following form \cite{Jackson}: 
\beqa\label{1}
\Evec_{\pvec}(\rvec)=-\frac{1}{3\eps_0}\,\pvec\,\delta(\rvec)+\frac{1}{4\pi\eps_0}\frac{3(\nuvec\cdot\pvec)\nuvec-\pvec}{r^3}=-\frac{1}{3\eps_0}\,\pvec\,\delta(\rvec) + \big(\Evec\big)_{r\ne 0}
\eeqa
where $\nuvec=\rvec/r$, and
\beqa\label{2}
\Bvec_{\mvec}(\rvec)= \frac{2\mu_0}{3}\,\mvec\,\delta(\rvec)+\frac{\mu_0}{4\pi}\frac{3(\nuvec\cdot\mvec)\nuvec-\mvec}{r^3}=
\frac{2\mu_0}{3}\,\mvec\,\delta(\rvec)+ \big(\Bvec\big)_{r\ne 0}
\eeqa
In these equations, by $(\dots)_{r\ne 0}$ we understand an expression in which the derivatives are calculated supposing $r\ne 0$, representing some well-known expressions of the fields.
The expressions from equations \eref{1} and \eref{2} are introduced in Ref.\  \cite{Jackson} as  conditions of compatibility with the average value of the  electric or magnetic field  over a spherical domain containing all the charges or currents inside. Another procedure for introducing equations \eref{1} and \eref{2} is based on an extension of the derivative $\d_i\d_j/(1/r)$ to the entire space \cite{Frahm}:
\beqa\label{3}
\d_i\d_j\frac{1}{r}=\,-\frac{4\pi}{3}\,\delta_{ij}\,\delta(\rvec)\,+\,\frac{3\,\nu_i\nu_j\,-\,\delta_{ij}}{r^3}.
\eeqa
A more pedagogical and suitable approach for understanding the origin of the difference between  the electric and magnetic cases,  is done in Ref.\  \cite{Leung06}.  Refs.\  \cite{Werner} and \cite{Leung07}  contain  generalizations of the equations \eref{1} and \eref{2} to the dynamic case for  oscillating electric and magnetic dipoles.
\par The objective of the present paper is to establish the singularities of the electromagnetic field associated to a system of electric charges and currents assimilated with a point-like multipolar system. These singularities are established for an arbitrary multipolar order for both, the static and the dynamic cases.
\par In section \ref{delta}, the procedure of separating the $\delta$-type singularities of the wave equation retarded solution  is described. In the next section, based on the exterior solution of a given static electric charge and current distribution expressed in terms of  electric and magnetic multipoles of this distribution, i.e.\  the multipole expansion of the electromagnetic field,  we separate the $\delta$-type singularities of the fields $\Evec$ and $\Bvec$. The calculation is based on the invariance of the static electromagnetic field to the substitution of the Cartesian tensors representing the multipole electric and magnetic moments by the corresponding symmetric trace free ({\bf STF}) projections. In section \ref{dynamic}, the singularities in the dynamic case are established by employing also the invariance of the electromagnetic field to the substitution of multipole Cartesian moments by {\bf STF} tensors, but this time, generally different from the corresponding {\bf STF} projections. 
\par In Appendix A, we give some furmulae for the lower order spatial derivatives of the spherical wave retarded solution  used in the current exposition. In the second appendix, the reduction to the {\bf STF} tensors in the dynamic case and for low orders is presented.
\par We point out that the formalism presented in this issue has as mathematical basis the properties of the irreducible tensorial  representations of the proper rotations group \cite{Damour}. Although, for pedagogical  and larger accessibility reasons, we give an explicit calculation based on the properties of the tensor contractions such that the procedure has a simple  algebraic character. A counterpart of the procedure used in the present paper could be represented by the technique of the spherical function expansions and the cited issues can be a basis for such an approach.

\section{Some delta-function identities}\label{delta}
The treatment of some delta-function identities in Ref.\  \cite{Frahm} can be easily generalized to obtain the necessary delta-function identities in the dynamic case. Let us consider the solution $f(t-r/c)/r$ of the wave equation. 
\beqa\label{4}
\left(\Delta-\frac{1}{c^2}\frac{\d^2}{\d t^2}\right)\,\frac{f(t-\frac{r}{c})}{r}\,=\,-4\pi\,f(t)\,\delta(\rvec).
\eeqa
In the points different from the origin  $O$, this function is a solution of the homogeneous wave equation. The multiple partial derivatives of this function are given, for $r\ne 0$, by the formula:
\beqa\label{5}
\left(\d_{i_1}\dots\d_{i_n}\frac{f(\tau)}{r}\right)_{r\ne 0}=\suml^n_{l=0}\frac{1}{c^{n-l}r^{l+1}}C^{(n,\,l)}_{i_1\dots i_n}\frac{\d^{n-l}f(\tau)}{\d t^{n-l}},
\eeqa
where $\tau\,=\,t-r/c$. The coefficients are symmetric in $i_1,\dots,i_n$ and can be expressed as
\beqa\label{6}
C^{(n,\,l)}_{i_1\dots i_n}=\suml^{[\frac{n}{2}]}_{k=0}D^{(n,\,l)}_k\,\delta_{\{i_1\,i_2}\dots \delta_{i_{2k-1}\,i_{2k}}
\nu_{2k+1}\dots\nu_{i_n\}}.
\eeqa
In the last equation, $[\beta]$ is the integer part of $\beta$ and $\nuvec=\rvec/r$. By 
$A_{\{i_1\dots i_n\}}$ 
we understand the sum over all the  permutations of the symbols $i_q$ giving distinct terms. For the objective of the present paper, the coefficients $D$ in equation \eref{6} can be calculated directly from the successive 
derivative operations. We give in appendix A the corresponding expressions for $n\le 4$. 

 From equation \eref{4}, it is obvious that we have to consider the derivatives of $f(\tau)/r$ extended over the entire space as  distributions (generalized functions) which contain singular distributions, as $\delta$-functions for example, as separate terms. Let be a function $F(\rvec,t)$ and suppose the existence of the integral of the product $F(\rvec,t)\,\phi(\rvec)$, with $\phi(\rvec)$ an arbitrary smooth function (a test function from the domain of the distributions),  
 on the spherical region $\dom_R$, with arbitrary radius $R$,  delimited by the spherical surface $\Sigma_R$ with the center  in $O$.
 \beqa\label{7}
 \int_{\dom_R}\rmd^3x\,F(\rvec,t)\,\phi(\rvec)=\lim_{\eps\to 0}\int_{\eps<r < R}\rmd^3x\,F(\rvec,t)\,\phi(\rvec)
 \eeqa
 The integral can be expressed excluding from the domain $\dom_R$ a spherical domain of radius $\eps$ centered in $O$. Writing this last limit of integrals, we can interpret the function $F(\rvec,t)$ as a distribution  defined by
 \beqan
 \left\langle\,\left(F(\rvec,t)\right)_{r\ne 0},\;\phi(\rvec)\right\rangle=\lim_{\eps\to 0}\int_{\dom_R\setminus\dom_\eps}\rmd^3x\,(F(\rvec,t))_{r\ne 0}\,\phi(\rvec)\ .
 \eeqan
 This distribution can be extended such that its support includes the point $O$. A new term $\theta(\eps-r)\,F(\rvec,t)$ can be naturally  introduced by the identity
 \beqan
 F_\eps(\rvec,t)=\theta(\eps-r)\,F(\rvec,t)+\theta(r-\eps)F(\rvec,t),
 \eeqan
 associated with the extension of the integral to the entire domain $\dom_R$:
 \beqa\label{8}
 \left\langle\,F(\rvec,t),\,\phi(\rvec)\right\rangle=\lim_{\eps\to 0}\left[\int_{\dom_\eps}\rmd^3x\,F(\rvec,t)\,\phi(\rvec)+\int_{\dom_R\setminus\dom_\eps}\rmd^3x\,F(\rvec,t)\,\phi(\rvec)\right].
 \eeqa
 Moreover, we suppose the existence of the integral \eref{7} for the partial derivatives of $F$.
  Let us consider the partial derivative $\d_iF(\rvec,t)$ and the problem of extending this function as a distribution. 
  The definition \eref{8} becomes:
\beqa\label{9}
 \left\langle\,F(\rvec,t),\,\phi(\rvec)\right\rangle&=&\lim_{\eps\to 0}\left[\oint_{\Sigma_\eps}\rmd S\,\nu_i\,F(\rvec,t)\phi(\rvec)-\int_{\dom_\eps}\rmd^3 x\,F(\rvec,t)\,\d_i\phi(\rvec)\right.\nonumber\\
&&\,\ \ +\left.\int_{\dom_R\setminus \dom_\eps}\,\rmd^3x\,\d_iF(\rvec,t)\,\phi(\rvec)\right],
\eeqa
where $\Sigma_{\eps}$ is the sphere of radius $\eps$ centered in $O$  and the Gauss theorem was employed.
Let us apply this definition to the derivative $\d_i\d_j(f(\tau)/r)$ and, for simplifying the notation, let 
\beqan
D_{i_1\dots i_n}(\rvec,t)=\d_{i_1}\dots \d_{i_n}\,\frac{f(\tau)}{r},
\eeqan
such that we can write
\beqa\label{10}
 \left(D_{ij},\,\phi\right)&=& \lim_{\eps\to 0}\left[\oint_{\Sigma_\eps}\,\rmd S\,\nu_i\,\d_j\frac{f(\tau)}{r}\,\phi(\rvec)- \int_{\dom_\eps}\,\rmd^3x\,\d_j\frac{f(\tau)}{r}\,\d_i\phi(\rvec)\right.\nonumber\\
&&\,\ \ + \left. \int_{\dom_R\setminus\dom_\eps}\rmd^3x\,\big(\d_i\d_j\frac{f(\tau)}{r}\big)\,\phi(\rvec)\right].              \eeqa
Since the last integral on the domain $\dom_R\setminus\dom_\eps$ represents the distributions associated with the $F$- expressions for $r\ne 0$ and, in  the case of the electromagnetic  field, will be the well-known expressions of the multipole expansions, in the following we consider only that part of $\left\langle D_{ij}\right\rangle$ containing singular distributions with point-like support i.e., actually,  the difference
\beqa\label{11}
\left\langle \left(D_{ij}\right)_{(0)},\;\phi\right\rangle&=&\left\langle D_{ij},\;\phi\right\rangle-\lim_{\eps\to 0}\int_{\dom_R\setminus\dom_\eps}\rmd^3x\,\big(\d_i\d_j\frac{f(\tau)}{r}\big)\,\phi(\rvec)
\nonumber\\
&=&\lim_{\eps\to 0}\left[\oint_{\Sigma_\eps}\,\rmd S\,\nu_i\,\d_j\frac{f(\tau)}{r}\,\phi(\rvec)- \int_{\dom_\eps}\,\rmd^3x\,\d_j\frac{f(\tau)}{r}\,\d_i\phi(\rvec)\right].
\eeqa
By $D_{(\rvec_0)}$ we denote a distribution having as support the point given by the vector $\rvec_0$.
In the first limit, we make use of equation \eref{5} and of the expression of the coefficients $C^{(1,l)}_i$ from Appendix A:
\beqan
\lim_{\eps\to 0}\oint_{\Sigma_\eps}\,\rmd S\,\nu_i\,\d_j\frac{f(\tau)}{r}\,\phi(\rvec)=-\lim_{\eps\to 0}\oint_{\Sigma_\eps}\,\rmd S\,\nu_i\nu_j\big[\frac{1}{cr}\dot{f}(\tau)+\frac{1}{r^2}f(\tau)\big]\,\phi(\rvec)\ .
\eeqan
Inserting the Taylor series of the function $\phi(\rvec)$ and since on the sphere $r=\eps$, we can write
\beqa\label{12}
\lim_{\eps\to 0}\oint_{\Sigma_\eps}\,\rmd S\nu_i\,\d_j\frac{f(\tau)}{r}\,\phi(\rvec) \nonumber\\
=-4\pi\lim_{\eps\to 0}\,\int\rmd\Omega(\nuvec)\,\nu_i\nu_j\left[\frac{\eps}{c}\,\dot{f}(\tau_\eps)  +f(\tau_\eps)\right] \left[\phi(0)+\eps\nu_k\left(\d_k\phi\right)_0+\dots\right]
\eeqa
where $\tau_\eps=t-\eps/c$.
Let us introduce the angular average:
\beqa\label{13}
\langle g(\nuvec)\rangle=\frac{1}{4\pi}\int\,g(\nuvec)\,\rmd\Omega(\nuvec)\ .
\eeqa
Particularly, we have the well-known  formula \cite{Thorne}:
\beqa\label{14}
\langle\nu_{i_1}\dots\nu_{i_n}\rangle=\left\{\begin{array}{c}0,\;\;\;\;\;\;\;\;\;\;\;\;\;\;\;\;\;\;\;\;\;\;\;\;\;\;n=2k+1,\\
\frac{1}{(n+1)!!}\,\delta_{\{i_1i_2}\dots\delta_{i_{n-1}i_n\}},\;\;\;\;\;\;\;\;\;\;n=2k,\;\;\;\;\;k=0,1,\dots\end{array}\right. 
\eeqa
  Excepting the term containing the product $f(\tau_\eps)\,\phi(0)$, all the terms in equation \eref{12} are proportional to positive powers of $\eps$ and, consequently, vanish with $\eps\to 0$ such that
 \beqan
 \lim_{\eps\to 0}\oint_{\Sigma_\eps}\,\rmd S\,\nu_i\,\d_j\frac{f(\tau)}{r}\,\phi(\rvec)&=&-4\pi\left\langle\,\nu_i\nu_j\right\rangle\,f(t)\,\phi(0)\\
&=&-\frac{4\pi}{3}\delta_{ij}f(t)\phi(0)=-\frac{4\pi}{3}\delta_{ij}f(t)\left\langle\delta(\rvec),\;\phi(\rvec)\right\rangle\ .
 \eeqan
  Considering the second integral in the right-hand side of equation \eref{11}, we can write
\beqa\label{15}
\fl&& \lim_{\eps\to 0}\int_{\dom_\eps}\rmd^3x\,\,\d_j\frac{f(\tau)}{r}\,\d_i\phi(\rvec)\nonumber\\
\fl\;\;\;\;\;&=&-\lim_{\eps\to 0}\int^\eps_0r^2\,\rmd    r\int\rmd\Omega(\nuvec)\,\nu_j\big[\frac{1}{cr}\dot{f}(\tau)+\frac{1}{r^2}f(\tau)\big]\big[(\d_i\phi)_0+r\nu_k(\d_i\d_k\phi)_0+\dots\big]=0
\eeqa
such that finally
\beqa\label{16}
\left(\d_i\d_j\frac{f(\tau)}{r}\right)_{(0)}=-\frac{4\pi}{3}\delta_{ij}f(t)\,\delta(\rvec)\ .                   \eeqa      
  
 In the static case,  $f(t)=1$, equation \eref{16} becomes  equation (3) from Ref.\  \cite{Frahm} after adding the regular distribution represented by the last integral from equation \eref{10}.
 \par Let us consider the distribution $D_{ijk}$. Considering only the part having $O$ as support,
\beqa\label{17}
\fl \;\;\;\;\;\;\;\;\;\;\;\;\left(\left(D_{ijk}\right)_{(0)},\,\phi\right)
=\lim_{\eps\to 0}\left[\oint_{\Sigma_\eps}\rmd S\,\nu_i\,\d_j\d_k\frac{f(\tau)}{r}\phi(\rvec)-
\int_{\dom_\eps}\rmd^3x\,\big(\d_j\d_k\frac{f(\tau)}{r}\big)\,\d_i\phi(\rvec)\right].
\eeqa
 The surface integral becomes:
 \beqan
 \fl \;\;\;\;\;\;\;\lim_{\eps\to 0}\oint_{\Sigma_\eps}\rmd S\,\nu_i\,\d_j\d_k\frac{f(\tau)}{r}\,\phi(\rvec)&=&\lim_{\eps\to 0}\oint_{\Sigma_\eps}\rmd S\,\nu_i\big[\frac{1}{c^2r}\nu_j\nu_k\,\ddot{f}(\tau)+\frac{1}{cr^2}(3\nu_j\nu_k-\delta_{jk})\dot{f}(\tau)\nonumber\\
 \fl&&\ \ \ +\frac{1}{r^3}(3\nu_j\nu_k-\delta_{jk})f(\tau)\big]\phi(\rvec)
 \eeqan
 and, introducing the Taylor series for  $\phi(\rvec)$,
\beqa\label{18}
 \fl &\lim_{\eps\to 0}&\oint_{\Sigma_\eps}\rmd S\,\nu_i\,\d_j\d_k\frac{f(\tau)}{r}\,\phi(\rvec)=
 4\pi\,\lim_{\eps\to 0}\left\langle\,\left[\frac{\eps}{c^2}\nu_i\nu_j\nu_k\,\ddot{f}(\tau_\eps)
 \right.\right.\nonumber\\
\fl&+&\left.\left.\frac{1}{c}(3\nu_i\nu_j\nu_k-\nu_{i}\delta_{jk}\dot{f}(\tau_\eps)+\frac{1}{\eps}(3\nu_i\nu_j\nu_k-\nu_{i}\delta_{jk})f(\tau_\eps)\right]\, 
\left[\phi(0)+\eps\nu_l\left(\d_l\phi\right)_0+\dots\right]\right\rangle\, .
 \eeqa
All the terms containing the products $\ddot{f}(\tau_\eps)\phi(0)$ and $\dot{f}(\tau_\eps)\phi(0)$ give null results being multiplied by averages of three factors $\nu$. The other terms containing the factors $\ddot{f}(\tau_\eps)$ and $\dot{f}(\tau_\eps)$ give also null results being proportional to positive powers of $\eps$.
The term provided by the product $f(\tau_\eps)\phi(0)$ contains the factor $\eps^{-1}$, but it has also a null limit since $\langle 3\nu_i\nu_j\nu_k-\nu_{i}\delta_{jk}\rangle=0$. Only the  product $f(\tau_\eps)\,\left(\d_l\phi\right)_0$ has a factor independent of $\eps$ having a limit different from zero. 
\par Therefore, 
\beqa\label{19}
\lim_{\eps\to 0}\oint_{\Sigma_\eps}\rmd S\,\nu_i\,\d_j\d_k\frac{f(\tau)}{r}\,\phi(\rvec)&=&4\pi(3\nu_i\nu_j\nu_k\nu_l-\nu_i\nu_l\delta_{jk})f(t)\left(\d_l\phi\right)_0\nonumber\\
&=&4\pi\left(\frac{1}{5}\delta_{\{ij}\delta_{kl\}}-\frac{1}{3}\delta_{il}\delta_{jk}\right)\,f(t)\,\left(\d_l\phi\right)_0
\eeqa

 Concerning the integral on $\dom_\eps$ from equation \eref{17}, we have to observe that, beginning from this derivative order, there is a non-zero contribution for $\eps\to 0$ \cite{Frahm}. Indeed, introducing equation \eref{16} in equation \eref{17} and observing that the term $(\d_j\d_k(f(\tau)/r))_{r\ne 0}$ gives a null contribution to the limit for $\eps\to 0$, we can write
\beqa\label{20}
\fl -\lim_{\eps\to 0}\int_{\dom_\eps}\rmd^3x\,\d_j\d_k\frac{f(\tau)}{r}\,\,\d_i\phi(\rvec)=\frac{4\pi}{3}\int_{\dom_\eps}\rmd^3x\,f(t)\delta_{jk}\delta(\rvec)\d_i\phi(\rvec)=\frac{4\pi}{3}\,f(t)\delta_{jk}(\d_i\phi)_0\ .
\eeqa
Finally, equations \eref{17}, \eref{18} and \eref{19} give 
\beqa\label{21}
(D_{ijk})_{(0)}=-\frac{4\pi}{5}f(t)\,\delta_{\{ij}\d_{k\}}\delta(\rvec)\ .
\eeqa
For the static  case, $f(t)=1$; this result corresponds to equation (4) from Ref.\  \cite{Frahm}.
\par Obviously, this procedure becomes very complicated for higher order derivatives. Fortunately, for the electromagnetic field some invariance properties allow a considerable simplification of such calculations.

\section{Singularities of the electromagnetic field: the static case}\label{static}
\par Let us consider the multipole expansions of the electrostatic and magnetostatic fields. Given an electric charge and current distribution with support included in the domain $\dom$,   the scalar potential is expressed  in the exterior of this domain by the following multipolar series:
 \beqa\label{22}
\Phi(\rvec)=\frac{1}{4\pi\eps_0}\suml_{n\ge 0}\frac{(-1)^n}{n!}\d_{i_1}\dots\d_{i_n}\frac{\psf_{i_1\dots i_n}}{r}
=\frac{1}{4\pi\eps_0}\suml_{n\ge 0}\frac{(-1)^n}{n!}\nablav^n\vert\vert\frac{\psft^{(n)}}{r}\ .
\eeqa
In this expansion, the coordinate system origin $O$ is supposed in $\dom$ and
 $\psft^{(n)}$ is  the $n$-th order electric multipolar moment defined by the Cartesian components in the general dynamic case:
\beqa\label{23}
\psf_{i_1\dots i_n}(t)=\int_{\dom}\,\rmd^3x\,\,x_{i_1}\dots x_{i_n}\,\rho(\rvec,t):\;\;\psft^{(n)}(t)=\intl_{\dom}\rmd^3x\,\rvec^n\,\rho(\rvec,t).
\eeqa
 In equation \eref{22}  we employed the following notation for tensorial contractions:
\beqa\label{24}
 ({\Atens}^{(n)}||{\Btens}^{(m)})_{i_1 \cdots i_{|n-m|}}
=\left\{\begin{array}{ll}
A_{i_1 \cdots i_{n-m}j_1 \cdots j_m}B_{j_1 \cdots j_m} & ,\; n>m\\
A_{j_1 \cdots j_n}B_{j_1 \cdots j_n} & ,\; n=m\\
A_{j_1 \cdots j_n}B_{j_1 \cdots j_n i_1 \cdots i_{m-n}} & ,\; n<m
\end{array} \right..
\eeqa
For the vector potential in the exterior of the domain $\dom$, we have
 \beqa\label{25}
\Avec(\rvec)&=&\frac{\mu_0}{4\pi}\suml_{n\ge 1}\frac{(-1)^{n-1}}{n!}\nablav\times\left(\nablav^{n-1}\vert\vert\msft^{(n)}\right)\nonumber\\
&=&\frac{\mu_0}{4\pi}\evec_i\eps_{ijk}\d_j\suml_{n\ge 1}\frac{(-1)^{n-1}}{n!}\d_{i_1}\dots\d_{i_{n-1}}\frac{\msf_{i_1\dots i_{n-1}\,k}}{r},
\eeqa
where $\msft^{(n)}$ is the magnetic $n$-th order moment defined by the Cartesian components  \cite{Castell}:
\beqa\label{26}
 \msf_{i_1\dots i_n}(t)=\frac{n}{n+1}\int_{\dom}\rmd^3x\,\,x_{i_1}\dots x_{i_{n-1}}\big(\rvec\times\jvec(\rvec,t)\big)_{i_n},
 \eeqa
 or, with tensorial notation:
 \beqan
 \msft^{(n)}(t)=\frac{n}{n+1}\intl_{\dom}\rmd^3x\,\,\rvec^n\times\jvec(\rvec,t)\ .
\eeqan
For the  multipole  expansion of the electric field $\Evec(\rvec)=-\nablav\,\Phi(\rvec)$, equation \eref{22} becomes:
 \beqa\label{27}
\fl\;\;\;\;\;\Evec(\rvec)=\frac{1}{4\pi\eps_0}\suml_{n\ge 1}\frac{(-1)^{n-1}}{n!}\nablav^{n+1}\vert\vert\frac{\psft^{(n)}}{r}=
\frac{1}{4\pi\eps_0}\evec_i\suml_{n\ge 1}\frac{(-1)^{n-1}}{n!}\d_i\,\d_{i_1}\dots\d_{i_n}\frac{\psf_{i_1\dots i_n}}{r},
\eeqa
where, for simplicity, the electric charged system  is considered neutral ($Q=0$).
\par The corresponding expansion of the magnetic field $\Bvec(\rvec)=\nablav\times\Avec(\rvec)$ is given by
 \beqa\label{28}
\fl \;\;\;\;\;\;\;\;\;\;\;\;\;\;\;\;\;\;\;\;\;\;\Bvec(\rvec)&=&\frac{\mu_0}{4\pi}\nablav\times\suml_{n\ge 1}\frac{(-1)^{n-1}}{n!}\nablav\times\left(\nablav^{n-1}\vert\vert\frac{\msft^{(n)}}{r}\right)\nonumber\\
\fl&=&\frac{\mu_0}{4\pi}\suml_{n\ge 1}\frac{(-1)^{n-1}}{n!}\left[\nablav\cdot\left(\nablav^n\vert\vert\frac{\msft^{n)}}{r}\right)
-\Delta\left(\nablav^{n-1}\vert\vert\frac{\msft^{(n)}}{r}\right)\right]\ .
 \eeqa
 We have to search the singularities of $\Evec$ and $\Bvec$ given by equations \eref{27} and \eref{28}. It appears that  cumbersome calculations are involved for higher $n$ if we apply the formulae for derivatives of an arbitrary function $f(\tau)/r$ as in the previous section. However, we can employ an invariance property of the electromagnetic static field to the substitutions of all moments $\psft^{(n)}$ and $\msft^{(n)}$ for all $n$ by their corresponding symmetric and trace-free {\bf STF} projections $\pct^{(n)}$ and $\mct^{(n)}$ \cite{cv-sc,Gonzales,cv02}.
  Retaining the notation $\pvec$ and $\mvec$ for  the first order moments, this invariance stands  for the invariance of the electromagnetic field, in the static case, to the following substitutions:
 \beqa\label{29}
\pvec,\,\psft^{(2)},\,\psft^{(3)},\,\dots\,&\to&\,\pvec,\,\pct^{(2)},\,\pct^{(3)}\dots~;\nonumber\\
\mvec,\,\msft^{(2)},\,\msft^{(3)},\,\dots\,&\to&\,\mvec,\,\mct^{(2)},\,\mct^{(3)}\dots\ .
\eeqa
 These {\bf STF} tensors can be expressed by the following formulae:
 \beqa\label{30}
 \pc_{i_1\dots      i_n}&=&\frac{(-1)^n}{(2n-1)!!}\int_{\dom}\,\rmd^3x\,\rho(\rvec)\rvec^{2n+1}\d_{i_1}\dots \d_{i_n}\frac{1}{r},\nonumber\\
 \mc_{i_1\dots i_n}&=&\frac{(-1)^{n+1}}{(n+1)(2n-1)!!}\suml^n_{\la=1}\int_{\dom}\rmd^3x\,\rvec^{2n+1}\left(\jvec\times\nablav\right)_{i_{\la}}\d^{(\la)}_{i_1\dots i_n}\frac{1}{r},
  \eeqa
 which, actually, differ from the projections by  numerical factors. Here, 
 \beqan
 \d^{(\la)}_{i_1\dots i_n}=\d_{i_1}\dots \d_{i_{\la-1}}\d_{i_{\la+1}}\dots \d_{i_n}\ .
 \eeqan
 \par Let us consider the delta-singularity corresponding to the electric dipolar field:
 \beqa\label{31}
 \Evec^{(1)}_{(0)}=\frac{1}{4\pi\eps_0}\evec_i\d_i\d_j\frac{p_j}{r}
=-\frac{1}{3\eps_0}\pvec\,\delta(\rvec),
 \eeqa
 a result obtained applying directly equation \eref{16} for $f(t)=p_j$. Concerning the 4-polar term from $\Evec$, we consider firstly the expansion \eref{27} expressed by the primitive moments $\psft^{(n)}$:
 \beqa\label{32}
\Evec^{(2)}(\rvec)=-\frac{1}{8\pi\eps_0}\evec_i\d_i\d_j\d_k\frac{\psf_{jk}}{r}
\eeqa
and from equation \eref{21}:
 \beqa\label{33}
\fl\Evec^{(2)}_{(0)}&=&\frac{1}{10\eps_0}\evec_i\,\psf_{jk}\delta_{\{ij}\d_{k\}}\delta(\rvec)
=\frac{1}{10\eps_0}\evec_i\,\psf_{jk}\left(\delta_{ij}\d_k\delta(\rvec)+\delta_{ik}\d_j\delta(\rvec)+\delta_{jk}\d_i\delta(\rvec)\right)\nonumber\\
\fl &=&\frac{1}{10\eps_0}\evec_i\left(2\psf_{ij}\d_j\delta(\rvec)+\psf_{jj}\d_i\delta(\rvec)\right)=\frac{1}{5\eps_0}\evec_i\left(\psf_{ij}\d_j\delta(\rvec)+\frac{1}{2}\psf_{jj}\d_i\delta(\rvec)\right).
\eeqa
Note that $\psft^{(2)}$ is symmetric.
 \par The substitution $\psf_{ij}\,\to\,\pc_{ij}$ in equation \eref{33} gives
 \beqa\label{34}
\Evec^{(2)}_{(0)}=\frac{1}{5\eps_0}\evec_i\pc_{ij}\d_j\delta(\rvec)=\frac{1}{5\eps_0}
\pct^{(2)}\vert\vert\nablav\delta(\rvec),
\eeqa
since $\pct^{(2)}$ is trace-less ($\pc_{jj}=0$).
\par From this simple example, it becomes obvious that it is more simple providing formulae for the delta-singularities directly for {\bf STF} tensors. Let us consider the 4-pole approximation using the definition
 \beqa\label{35}
\fl\left\langle\,\Evec^{(2)}_{(0)},\,\phi\right\rangle&=&-\frac{1}{8\pi\eps_0}\lim_{\eps\to 0}\evec_i\int_{\dom_\eps}\rmd^3x\,\left(\d_i\d_j\d_k\frac{\pc_{jk}}{r}\right)\,\phi(\rvec)\nonumber\\
\fl&=&-\frac{1}{8\pi\eps_0}\evec_i\lim_{\eps\to 0}\left[\oint_{\Sigma_\eps}\rmd S\,\pc_{jk}\nu_i\,\left(\d_j\d_k\frac{1}{r}\right)\,\phi(\rvec)-\int_{\dom_\eps}\rmd^3x\,\pc_{jk}\d_j\d_k\frac{1}{r}\;\d_i\phi(\rvec)\right].
\eeqa
In the last integral over $\dom_\eps$ from equation \eref{35}, we can introduce equation \eref{16} obtaining
\beqan
\fl&\lim_{\eps\to 0}&\int_{\dom_\eps}\rmd^3x\,\pc_{jk}\left[-\frac{4\pi}{3}\delta_{jk}\delta(\rvec)+\left(\d_j\d_k\frac{1}{r}\right)_{r\ne 0}\right]\d_i\phi(\rvec)=\lim_{\eps\to0}\int_{\dom_\eps}\rmd^3x\,\pc_{jk}\left(\d_j\d_k\frac{1}{r}\right)_{r\ne 0}\,\d_i\phi(\rvec)\nonumber\\
\fl&=&\lim_{\eps\to0}\int_{\dom_\eps}\rmd^3x\,\pc_{jk}\frac{1}{r^3}\left(3\nu_j\nu_k-\delta_{jk}\right)\d_i\phi(\rvec)
=3\lim_{\eps\to0}\int_{\dom_\eps}\rmd^3x\,\frac{1}{r^3}\pc_{jk}\,\nu_j\nu_k\,\d_i\phi(\rvec)\nonumber\\
\fl&=&3\lim_{\eps\to 0}\int^\eps_0\frac{\rmd r}{r}\oint\rmd\Omega\,\pc_{jk}\,\nu_j\nu_k\,\left[(\d_i\phi)_0+r\nu_l(\d_i\d_l\phi)_0+\dots\right]=0,
\eeqan
since the term $\pc_{jk}\nu_j\nu_k(\d_i\phi)_0$ provides the expression $\pc_{jk}\delta_{jk}=0$ and for the other terms from the $\d_i\phi$ series has null limits containing, after integration over $r$, positive powers of $\eps$. The same result can be obtained by a repetition of the integration by parts in the last volume integral. Therefore, we can write
\beqan
\fl\left\langle\,\Evec^{(2)}_{(0)},\,\phi\right\rangle&=&-\frac{1}{8\pi\eps_0}\evec_i\lim_{\eps\to 0}\oint_{\Sigma_\eps}\rmd S\,\pc_{jk}\nu_i\,\left(\d_j\d_k\frac{1}{r}\right)\,\phi(\rvec)\nonumber\\
\fl&=&-\frac{1}{2\eps_0}\evec_i\lim_{\eps\to 0}\pc_{jk}\frac{1}{\eps}\left\langle\nu_i(3\nu_j\nu_k-\delta_{jk})\left(\phi(0)+\eps\nu_l(\d_l\phi)_0+\dots)\right)\right\rangle\nonumber\\
\fl&=&-\frac{1}{2\eps_0}\evec_i\pc_{jk}\left\langle\,3\nu_i\nu_j\nu_k\nu_l\right\rangle\left(\d_l\phi)\right)_0=-\frac{1}{10\eps_0}\evec_i\pc_{jk}\left(\delta_{\{ij}\delta_{kl\}}\right)(\d_l\phi)_0\nonumber\\
\fl&=&-\frac{1}{5\eps_0}\pct^{(2)}\vert\vert\left(\nablav\phi\right)_0,
\eeqan
i.e.\   the result \eref{34}.
\par Let us consider the general terms from the series representing the electric and  magnetic fields in equations \eref{27} and \eref{28} with the substitutions \eref{29} performed. Denoting by $\Tsft^{(n)}$ an $n$-th order {\bf STF} tensor, we observe the presence in both equations of an expression of the form 
\beqan
\nablav\left(\nablav^n\vert\vert\frac{\Tsf^{(n)}}{r}\right)=\evec_i\d_i\,\d_{i_1}\dots \d_{i_n}\frac{\Tsf_{i_1\dots i_n}}{r}\ .
\eeqan
Searching the extension of this function to the entire space, we introduce the distribution $\bbox{I}(\Tsf^{(n)})$ by 
 \beqa\label{36a}
 \left\langle\,\bbox{I}(\Tsf^{(n)}),\;\phi\right\rangle=\lim_{\eps\to 0}\int_{\dom_\eps}\rmd^3x\,\nablav\left(\nablav^n\,\frac{\Tsft^{(n)}}{r}\right)\ .
 \eeqa
 Writing the last equation as
 \beqa\label{36b}
 \fl\;\;\; \left\langle\,\bbox{I}(\Tsf^{(n)}),\;\phi\right\rangle=\lim_{\eps\to 0}\left[\oint_{\Sigma_\eps}\rmd S\,\nuvec\left(\nablav^n\vert\vert\frac{\Tsft^{(n)}}{r}\right)\,\phi(\rvec)-\int_{\dom_\eps}\rmd^3x\,\left(\nablav^n\vert\vert\frac{\Tsft^{(n)}}{r}\right)\,\nablav\phi(\rvec) \right],
\eeqa 
 we can consider separately the limit of the surface integral:
  \beqan
 \left\langle\,\bbox{I}_\sigma(\Tsft^{(n)}),\,\phi\right\rangle&=&\lim_{\eps\to 0}\oint_{\Sigma_\eps}\rmd S\,\nuvec\left(\nablav^n\vert\vert\frac{\Tsft^{(n)}}{r}\right)\,\phi(\rvec)\\
 &=&\evec_i\,\lim_{\eps\to 0}\oint_{\Sigma_\eps}\rmd S\,\nu_i\Tsf_{i_1\dots i_n}\left(\d_{i_1}\dots\d_{i_n}\frac{1}{r}\right)\,\phi(\rvec).
  \eeqan
 Applying equation \eref{5} and the Taylor series of $\phi(\rvec)$,
 \beqa\label{37a} 
  \left\langle\,\bbox{I}_\sigma(\Tsft^{(n)}),\,\phi\right\rangle 
 &=&4\pi \evec_i\lim_{\eps\to 0}\suml^\infty_{\al=0}\frac{\eps^{\al-n+1}}{\al!}\Tsf_{i_1\dots i_n}\left\langle \nu_i\,C^{(n,n)}_{i_1\dots i_n}\,\nu_{i_{n+1}}\dots \nu_{i_{n+\al}}\right\rangle \nonumber\\
 & & \ \ \ \times \left(\d_{i_{n+1}}\dots \d_{i_{n+\al}}\,\phi\right)_0.
 \eeqa
 Let us evaluate the tensorial contraction which is present in the general term from the previous series:
 \beqa\label{37b} 
 \Tsf_{i_1\dots i_n}\left\langle\,\nu_i\,C^{(n,n)}_{i_1\dots i_n}\nu_{i_{n+1}}\dots\nu_{i_{n+\al}}\right\rangle\ .
\eeqa
 From equation \eref{6}, we can easily see that all the terms containing at least a symbol $\delta_{i_qi_s}$ with $1\le q,\,s\,\le n$  give null results by contraction with $\Tsft^{(n)}$. An exception is represented by the term containing the factor $D^{(n,n)}_0\,\nu_{i_1}\dots \nu_{i_n}$ which, together with the factor $\nu_{i_{n+1}}\dots \nu_{i_{n+\al}}\,\nu_i$, can give results different from zero. Since
 \beqa\label{37c} 
 D^{(n,n)}_0=(-1)^n\,(2n-1)!!\ ,
\eeqa
 the series \eref{37a} can be written as
 \beqa\label{38a} 
\fl\;\;\;\;\;\;\;\;\left\langle\,\bbox{I}_\sigma(\Tsft^{(n)}),\,\phi\right\rangle&=&4\pi\evec_i(-1)^n(2n-1)!!\;\lim_{\eps\to 0}\suml^\infty_{\al=0}\frac{\eps^{\al-n+1}}{\al!}\nonumber\\
 \fl & \times& \Tsf_{i_1\dots i_n}\left\langle\,\nu_{i_1}\dots \nu_{i_n}\,\nu_{i_{n+1}}\dots \nu_{i_{n+\al}}\,\nu_i\right\rangle\,\left(\d_{i_{n+1}}\dots\d_{i_{n+\al}}\,\phi\right)_0\ .
  \eeqa
Firstly, we have to see what happens for negative powers of $\eps$ in the terms of the series from the last equation. These negative powers  are factors in the terms verifying the inequality $\al\,<\,n-1$. The expression  
 is different from zero  if and only if it contains some terms in which all the indices $i_1\dots i_n$ are present as  index pairs  of  Kronecker symbols $\delta_{i_ji_k}$ or $\delta_{i\,i_q}$ with $j,\,q=1,\,\dots, n$ and 
 $k\in\{n+1,\,\dots,\,n+\al\}$.  Otherwise, one obtains null values because of the contraction with an {\bf STF} tensor.   For $\al\,<\,n-1$  such terms cannot exist. 
 \par For $\al\,>\,n-1$, the corresponding terms from the series in equation \eref{38a} contain positive powers of $\eps$. Consequently, the corresponding limits for $\eps\,\to 0$  vanish. From this series only the term for which 
 \beqa\label{42}
 \al=n-1
 \eeqa
 can be different from zero and the result of the limit in equation \eref{38a} is given by
 \beqa\label{42a}
\fl\;\;\;\;\;\;\;\;\left\langle\,\bbox{I}_\sigma(\Tsft^{(n)}),\,\phi\right\rangle&=&\frac{4\pi(-1)^n(2n-1)!!}{(n-1)!}\nonumber\\
&&\ \ \ \times\evec_i\,\Tsf_{i_1\dots i_n}\left\langle\,\nu_{i_1}\dots \nu_{i_n}\,\nu_{i_{n+1}}\dots \nu_{i_{2n-1}}\,\nu_i\right\rangle\,\left(\d_{i_{n+1}}\dots \d_{i_{2n-1}}\,\phi\right)_0\nonumber\\
&=&\frac{4\pi(-1)^n(2n-1)!!}{(n-1)!}\left\langle\,\Tsft^{(n)}\vert\vert\nuvec^{2n}\right\rangle\vert\vert\left(\nablav^{n-1}\phi\right)_0\ .
 \eeqa
 The corresponding $\delta$-function  is therefore
 \beqa\label{43}
\bbox{I}_\sigma(\Tsft^{(n)})=-\frac{4\pi(2n-1)!!}{(n-1)!}\,\left\langle\Tsft^{(n)}\vert\vert\nuvec^{2n}
\right\rangle\vert\vert\nablav^{n-1}\delta(\rvec)\ .
 \eeqa
 Let us consider the contraction 
 \beqan
 \left\langle\,\Tsft^{(n)}\vert\vert\nuvec^{2n}\right\rangle=\Tsf_{i_1\dots i_n}\,\left\langle\, \nu_{i_1}\dots\nu_{i_n}\nu_{j_1}\dots\nu_{j_n}\right\rangle\ .
 \eeqan
 To this contraction contribute only the terms from the average of the $\nu$-product not containing factors $\delta_{i_ki_l}$, with $1\le k,\,l\le n$. According to equation \eref{14}, the terms giving non-zero contributions are of the form
 \beqan
 \frac{1}{(2n+1)!!}\,\delta_{i_1j_1}\dots \delta_{i_nj_n}
   \eeqan
  and all such terms are obtained considering  the $n!$ permutations of the indices $j_1\dots j_n$ in this product. Therefore, the final expression in equation \eref{43} is given by
 \beqa\label{44}
\bbox{I}_\sigma(\Tsft^{(n)})=-\,\frac{4\pi\,n}{ 2n+1}\,\Tsft^{(n)}\vert\vert\nablav^{n-1}\delta(\rvec)\ .
\eeqa
 \par Let us evaluate now the volume integral in equation \eref{36b}:
 \beqa\label{45}
 \fl\;\;\;\;\;\;&&\left\langle\,J(\Tsft^{(n)}),\;\nablav\phi\right\rangle=\lim_{\eps\to 0}\int_{\dom_\eps}\rmd^3x\,\left(\nablav^n\vert\vert\frac{\Tsft^{(n)}}{r}\right)\,\nablav\phi(\rvec)\nonumber\\
 \fl&&\ \ \ =\lim_{\eps\to 0}\left[\oint_{\Sigma_\eps}\rmd S\,\nuvec\vert\vert\left(\nablav^{n-1}\vert\vert\frac{\Tsft^{(n)}}{r}\right)\,\nablav\phi(\rvec)
 -\int_{\dom\eps}\rmd^3x\,\left(\nablav^{n-1}\vert\vert\frac{\Tsft^{(n)}}{r}\right)\,\nablav^2\phi(\rvec) \right]\ .
 \eeqa
 The part corresponding to the surface integral can be written as
 \beqa\label{46}
\left\langle\,J_\sigma(\Tsft^{(n)}),\;\nablav\phi\right\rangle=\evec_i\Tsf_{i_1\dots i_n}\lim_{\eps\to 0}\oint_{\Sigma_\eps}\rmd S\,\nu_{i_n}\,\left(\d_{i_1}\dots \d_{i_{n-1}}\frac{1}{r}\right)\,\d_i\phi(\rvec)\ .
 \eeqa
 Introducing the Taylor series for $\phi(\rvec)$ and standing out the average over $\nuvec$, we obtain
 \beqa\label{46a}
\fl \left\langle\,J_\sigma(\Tsft^{(n)}),\;\nablav\phi\right\rangle
&=&4\pi\,\evec_i\lim_{\eps\to 0}\suml^\infty_{\al=0}\frac{\eps^{\al-n+2}}{\al!}\Tsf_{i_1\dots i_n}\left\langle\,C^{(n-1,n-1)}_{i_1\dots i_{n-1}}\nu_{i_n}\dots \nu_{i_{n+\al}} \right\rangle\,\nonumber\\
&& \ \ \ \times \left(\d_{i_{n+1}}\dots \d_{i_{n+\al}}\d_i\phi\right)_0.
 \eeqa
Analogously to the reasoning from the previous case, we can see that the limit is zero since for $\al=n-2$ 
 \beqan
 \Tsf_{i_1\dots i_n}\left\langle\,C^{(n-1,n-1)}_{i_1\dots i_{n-1}}\nu_{i_n}\nu_{i_{n+1}}\dots \nu_{i_{2n-2}}
 \right\rangle=0.
 \eeqan
 Therefore, 
 \beqa\label{48}
 \left\langle\,J(\Tsft^{(n)}),\;\nablav\phi\right\rangle=-\lim_{\eps\to 0}\int_{\dom_\eps}\rmd^3x\,\left(\nablav^{n-1}\vert\vert\frac{\Tsft^{(n)}}{r}\right)\,\nablav^2\phi(\rvec)\ ,
\eeqa 
and by repeatedly applying the procedure, all terms cancel. Finally, only the surface integral from equation \eref{36b} gives a limit different from zero 
 and 
 \beqan
 \bbox{I}(\Tsft^{(n)})=\bbox{I}_\sigma(\Tsft^{(n)})\ .
 \eeqan
 
  Introducing the result \eref{44} in equation \eref{27}, we get
 \beqa\label{49}
 \Evec^{(n)}_{(0)}=\frac{(-1)^n}{(n-1)!\,(2n+1)\,\eps_0}\pct^{(n)}\vert\vert\nablav^{n-1}\,\delta(\rvec)\ .
\eeqa
 With this formula one obtains easily the results for $n=1$ and $n=2$. If we take the $3$-rd order term of the electric field, equation \eref{49} gets the form:
 \beqa\label{50}
\Evec^{(3)}_{(0)}=-\frac{1}{14\,\eps_0}\pct^{(3)}\vert\vert\nablav^2\,\delta(\rvec)=-\frac{1}{14\,\eps_0}\evec_i\pc_{ijk}\d_j\d_k\delta(\rvec)\ .
 \eeqa
 For the $4$-th order multipolar  term, equation \eref{49} reads
 \beqa\label{51}
 \Evec^{(4)}_{(0)}=\frac{1}{54\,\eps_0}\pct^{(4)}\vert\vert\nablav^3\,\delta(\rvec)
 \eeqa
 and so on.
 \par Concerning the magnetic field, selecting the singular point-like part of $\Bvec^{(n)}$ from equation \eref{28}, we can write
  \beqan
\fl\left\langle\,\Bvec^{(n)}_{(0)},\;\phi\right\rangle&=&\frac{(-1)^{n-1}\mu_0}{4\pi\,n!}\lim_{\eps\to 0}\intl_{\dom_\eps}\rmd^3x\,\mct^{(n)}\vert\vert\left[\nablav^{n+1}\frac{1}{r}
-\nablav^{n-1}\Delta\frac{1}{r}\right]\,\phi(\rvec)\nonumber\\
\fl&=&\frac{(-1)^{n-1}\mu_0}{4\pi\,n!}\lim_{\eps\to 0}\intl_{\dom_\eps}\rmd^3x\,\mct^{(n)}\vert\vert\left[\nablav^{n+1}\frac{1}{r}
+4\pi\left(\nablav^{n-1}\delta(\rvec)\right)\right]\,\phi(\rvec)\nonumber\\
&=&\frac{(-1)^{n-1}\mu_0}{4\pi\,n!}\lim_{\eps\to 0}\intl_{\dom_\eps}\rmd^3x\,\left(\mct^{(n)}\vert\vert\nablav^{n+1}\frac{1}{r}\right)\,\phi(\rvec)+\frac{\mu_0}{n!}\mct^{(n)}\vert\vert\left(\nablav^{n-1}\phi\right)_0,
\eeqan
since $\Delta(1/r)=-4\pi\delta(\rvec)$. Written explicitly, this last equation becomes:
 \beqa\label{53}
\left\langle\,\Bvec^{(n)}_{(0)},\;\phi\right\rangle&=&\frac{(-1)^{n-1}\,\mu_0}{4\pi\,n!}\,\evec_i\lim_{\eps\to 0}\int_{\dom_\eps}\rmd^3x\mc_{i_1\dots i_n}\d_i\left(\d_{i_1}\dots\d_{i_n}\frac{1}{r}\right)\phi(\rvec)\nonumber\\
&&\ \ \ +\frac{\mu_0}{n!}\evec_i\mc_{i_1\dots i_{n-1}\,i}\left(\d_{i_1}\dots \d_{i_{n-1}}\,\phi\right)_0\ .
 \eeqa
 The integral on $\dom_{\eps}$ from equation \eref{53} is of the same type as the integral  from equation \eref{36a} and we can apply the result \eref{44} obtained in the electric case, such that
 \beqan
\Bvec^{(n)}_{(0)}=\left(\frac{(-1)^n\,\mu_0}{(n-1)!(2n+1)}+\frac{(-1)^{n-1}\,\mu_0}{n!}\right)\mct^{(n)}\vert\vert\nablav^{n-1}\delta(\rvec),
 \eeqan
 i.e.\ 
 \beqa\label{53a}
 \Bvec^{(n)}_{(0)}=\frac{(-1)^{n-1}\,\mu_0\,(n+1)}{n!\,(2n+1)}\,\mct^{(n)}\vert\vert\nablav^{n-1}\,\delta(\rvec).
 \eeqa
 Let us apply equation \eref{53a}  for lower multipolar orders.
 \par For the dipolar case:
 \beqa\label{55}
 \Bvec^{(1)}_{(0)}=\frac{2\mu_0}{3}\mvec\,\delta(\rvec)\ .
 \eeqa
 For $n=2$, the quadrupolar case,
 \beqa\label{56}
\Bvec^{(2)}_{(0)}=-\frac{3\mu_0}{10}\mc^{(2)}\vert\vert\nablav\delta(\rvec)\ .
 \eeqa
 For $n=3$,
 \beqa\label{57}
\Bvec^{(3)}_{(0)}=\frac{2\mu_0}{21}\,\mct^{(3)}\vert\vert\nablav^2\delta(\rvec)\ .
\eeqa
\par Before passing to the problem of the variable electromagnetic field, we point out some  simple consequences of the formalism employed  in the present section concerning the  integrals over a spherical domain $\dom_R$  of the electrostatic and magnetostatic fields, supposing the support of electric charges or currents included in $\dom_R$.
\par Using equation \eref{22} for representing the electric potential on the spherical surface $\Sigma_R$, we can write, employing the invariance of $\Evec$ to the substitutions \eref{29}, \cite{Jackson}
\beqan
\fl\;\;\;\;\;\;\int_{\dom_R}\rmd^3x\,\Evec(\rvec)= -\oint_{\Sigma_R}\rmd S\,\nuvec \Phi(\rvec)  
=\frac{1}{4\pi\eps_0}\suml_{n\ge 1}\frac{(-1)^{n-1}}{n!}\oint_{\Sigma_R}\rmd S\,\,\nuvec\left(\nablav^n\vert\vert\frac{\pct^{(n)}}{r}\right).
\eeqan 
By applying the Gauss theorem in the above integral, there were supposed fulfilled all the conditions of continuity of the field on the surface $\Sigma$ of the domain $\dom$ where the electric charge is distributed.
\par Considering the surface integral corresponding to the $n$-th order,
\beqan
\fl\bbox{I}_R=\oint_{\Sigma_R}\rmd S\,\,\nuvec\left(\nablav^n\vert\vert\frac{\pct^{(n)}}{r}\right)=\evec_i\oint_{\Sigma_R}\rmd S\,\nu_i\,\d_{i_1}\dots \d_{i_n}\frac{\pc_{i_1\dots i_n}}{r}
=\frac{4\pi}{R^{n-1}}\evec_i\left\langle\,\nu_i\,C^{(n,n)}_{i_1\dots i_n}\right\rangle\,\pc_{i_1\dots i_n}.
\eeqan
 Excepting the case $n=1$, the contractions of the angular averages with the {\bf STF} tensor $\pct^{(n)}$ give null result such that
 \beqan
 \bbox{I}_R=4\pi\,\evec_i\left\langle\,\nu_i\nu_j\right\rangle\,p_j=-\frac{4\pi}{3}\,\pvec\ ,
  \eeqan
 and 
  \beqan
  \int_{\dom_R}\rmd^3x\,\Evec(\rvec)=-\frac{1}{3\eps_0}\,\pvec ,
  \eeqan 
  i.e.\  equation (4.18) from Ref.\  \cite{Jackson}.
  \par In the case of the magnetic field, using the exterior solution for $\Avec$ given by equation \eref{25}, we can write
  \beqan
\fl  \int_{\dom_R}\rmd^3x\,\Bvec(\rvec)=\oint_{\Sigma_r}\rmd S\,\nuvec\times \Avec=
   \frac{\mu_0}{4\pi}\suml_{n\ge 1}\frac{(-1)^{n-1}}{n!}\oint_{\Sigma_R}\rmd S\,\nuvec\times\left[\nablav\times\left(\nablav^{n-1}\vert\vert\frac{\mct^{(n)}}{r}\right)\right]\ .
  \eeqan
  Considering the surface integral corresponding to the $n$-th order,
  \beqan
  \bbox{J}_R&=&\evec_i\eps_{ijk}\eps_{klq}\oint_{\Sigma_r}\rmd S\,\nu_j\left(\d_l\d_{i_1}\dots \d_{i_n}\frac{1}{r}\right)\,\mc_{i_1\dots i_n}\\
&=&\evec_i\eps_{ijk}\eps_{klq}\oint_{\Sigma_r}\frac{\rmd \Omega(\nuvec)}{R^{n-1}}\nu_j\,C^{(n,n)}_{i_1\dots i_{n-1}\,l}
\mc_{i_1\dots i_{n-1}\,q}\\
&=&\frac{4\pi}{R^{n-1}}\evec_i\eps_{ijk}\eps_{klq}\left\langle\,\nu_jC^{(n,n)}_{i_1\dots i_{n-1}\,l}\right\rangle\mc_{i_1\dots i_{n-1}q}\ .
  \eeqan
 As in the electrostatic case, excepting the case $n=1$, the above contraction gives null result such that
 \beqan
 \bbox{J}_R=-4\pi\,\eps_{ijk}\eps_{klq}
\left\langle\,\nu_j\nu_l\right\rangle\,m_q=-\frac{4\pi}{3}\left(\delta_{il}\delta_{jq}-\delta_{iq}\delta_{jl}\right)\delta_{jl}m_q=\frac{8\pi}{3}\mvec
 \eeqan  
and consequently,
 \beqan
 \int_{\dom_R}\rmd^3x\,\Bvec(\rvec)=\frac{2\mu_0}{3}\;\mvec\ ,
 \eeqan 
  i.e.\  equation (5.62) from Ref.\  \cite{Jackson}.

 \section{Singularities of the electromagnetic field: the dynamic case}\label{dynamic}
In the dynamic case, the multipole expansions of the potentials in an arbitrary point exterior to the domain $\dom$ are given by the expressions \cite{cv02}: 
\beqa\label{58}
\Phi(\rvec,\,t)&=&\frac{1}{4\pi\eps_0}\suml_{n\ge 1}\frac{(-1)^n}{n!}\nablav\vert\vert\frac{\psft^{(n)}(\tau)}{r},\nonumber\\
\Avec(\rvec,\,t)&=&\frac{\mu_0}{4\pi}\suml_{n\ge 1}\frac{(-1)^{n-1}}{n!}\left[\nablav\times\big(\nablav^{n-1}\frac{\msft^{(n)}(\tau)}{r}\big)+\nablav^{n-1}\vert\vert\frac{\dot{\psft}^{(n)}(\tau)}{r}\right]\ .
\eeqa
As in the static case,  for simplicity, we consider a  neutral electric system $(Q=0)$. 
We write the multipole  expansion of the electric field $\Evec(\rvec,t)$ derived from the potentials given in equation \eref{58}:
\beqa\label{59}
\Evec(\rvec,\,t)&=&-\frac{1}{4\pi\eps_0}\suml_{n\ge 1}\frac{(-1)^n}{n!}\left\{\nablav\big(\nablav^n\vert\vert\frac{\psft^{(n)}(\tau)}{r}\big)-\frac{1}{c^2}\nablav^{n-1}\vert\vert\frac{\ddot{\psft}^{(n)}(\tau)}{r}\right.\nonumber\\
&&\ \ \ -\left.\frac{1}{c^2}\nablav\times\big[\nablav^{n-1}\vert\vert\frac{\dot{\msft}^{(n)}(\tau)}{r}\big]\right\}.
\eeqa
The multipolar expansion of the magnetic field is given by
\beqa\label{60}
\fl \;\;\;\;\;\Bvec(\rvec,t)&=&\frac{\mu_0}{4\pi}\nablav\times\suml_{n\ge 1}\frac{(-1)^{n-1}}{n!}\left[\nablav\times\big(\nablav^{n-1}\vert\vert\frac{\msft^{(n)}(\tau)}{r}\big)+\nablav^{n-1}\vert\vert\frac{\dot{\psft}^{(n)}(\tau)}{r}\right]\ .
\eeqa
Since
\beqan
\fl\;\;\;\;\nablav\times\left[\nablav\times\left(\nablav^{n-1}\vert\vert\frac{\msft^{(n)}(\tau)}{r}\right)\right]
=\nablav\left(\nablav^n\vert\vert\frac{\msft^{(n)}(\tau)}{r}\right)-\Delta\left(\nablav^{n-1}\vert\vert\frac{\msft^{(n)}(\tau)}{r}\right)\ ,
\eeqan
equation \eref{60} can be written as
\beqa\label{61}
\Bvec(\rvec,t)&=&\frac{\mu_0}{4\pi}\suml_{\ge 1}\frac{(-1)^{n-1}}{n!}\left[\nablav\left(\nablav^n\vert\vert\frac{\msft^{(n)}(\tau)}{r}\right)
-\Delta\left(\nablav^{n-1}\vert\vert\frac{\msft^{(n)}(\tau)}{r}\right)\right]\nonumber\\
&+&\frac{\mu_0}{4\pi}\suml_{n\ge 1}\frac{(-1)^{n-1}}{n!}\nablav\times\left(\nablav^{n-1}\vert\vert\frac{\dot{\psft}^{(n)}(\tau)}{r}\right)\ .
\eeqa
Equations \eref{59} and \eref{61} represent the fields $\Evec$ and $\Bvec$ for $r\ne 0$, where the multipole tensors considered at $t=\tau$ and divided by $r$ are solutions of the homogeneous wave equation. This property can be considered in any  processing of the field multipole expansions. The corresponding delta-singularities,  i.e.\  including the point $O$ in the domain of the field,  will be  expressed searching the extensions as generalized functions of the final expressions.
 \par As in the static case, for higher orders of multipolar terms, $n\ge 2$, the labor involved in the calculation becomes prohibitive if we operate with the primitive moments $\psft^{(n)}$ and $\msft^{(n)}$. Fortunately, there is  an invariance property of the electromagnetic field in the dynamic case, too, which allows us to replace in the expressions of $\Evec$ and $\Bvec$ the tensors $\psft^{(n)}$ and $\msft^{(n)}$  by symmetric and trace-free ({\bf STF}) tensors for any $n$:
\beqa\label{62} 
\psft^{(n)}\,\to\,\psftr^{(n)},\;\;\;\;\msft^{(n)}\,\to\,\msftr^{(n)}.
\eeqa
$\psftr^{(n)}$ and $\msftr^{(n)}$ are {\bf STF} tensors (see Refs.\  \cite{Dubovik-FEC, Dubovik-rep, RV} for $n\le 3$ and \cite{Damour,cv02,cv03} for arbitrary $n$). Theoretically, each such {\bf STF} tensor is an infinite series, but practically, we have to calculate only a finite number of terms from the multipole expansion such that the respective tensors are represented by finite sums. Consequently, we can operate in the following with the {\bf STF} tensors $: \pvr,\;\psftr^{(2)},\;\psftr^{(3)},\dots$ for the electric cases, 
and $\mvr,\;\msftr^{(2)},\dots$ for the magnetic ones. However, for not invoking  directly some results from the issues cited above, we introduce in  \ref{ap2} the lower orders {\bf STF} tensors employed in the present paper, together with the introduction of the {\bf STF} projections of the primitive momenta. In every case, we will establish what are the changes of different momentum tensors which compensate the effects of such substitutions.
\par Let us consider the problem of expressing the singularities  of the electromagnetic field in $O$ when we are interested only in the contribution of the electric dipolar moment $\pvec(t)$. In this case we have an elementary charged system characterized only by the dipolar electric moment as, for example, the harmonic oscillator. Another case is when we are interested only in the first multipole approximation for a complex system, as an atom.
 For $n=1$, considering the contributions of the electric  dipolar moment,  equations \eref{59} and \eref{16} give
\beqa\label{63}
 &&\Evec^{(1,p)}_{(0)}(\rvec,t)=-\frac{1}{3\eps_0}\pvec(t)\,\delta(\rvec)\ .
 \eeqa
The singularity in the case of a magnetic dipole is
\beqan
\fl\Bvec^{(1,m)}_{(0)}=\frac{\mu_0}{4\pi}\evec_i\left[\d_i\d_j\frac{m_j(\tau)}{r}-\evec_i\Delta\frac{m_i(\tau)}{r}\right]
=\mu_0\evec_i\left[-\frac{1}{3}m_j(t)\delta_{ij}\delta(\rvec)+m_i(t)\delta(\rvec)\right]\ ,
\eeqan 
i.e.\ 
\beqa\label{64}
\Bvec^{(1,m)}_{(0)}=\frac{2\mu_0}{3}\mvec(t)\delta(\rvec)\ .
 \eeqa
Comparing equations \eref{63} and \eref{64} with equations \eref{1} and \eref{2} from the static case, one observes  that the results for the dynamic case are obtained from  the static one by the substitutions $\pvec\,\to\,\pvec(t),\;\;\mvec\,\to\, \mvec(t)$.
 \par In equations \eref{59} and \eref{61}, for the multipole expansions of the fields $\Evec$ and $\Bvec$ we have a first general term of the form
 \beqan
\nablav^{n+1}\vert\vert\frac{\Tsft^{(n)}(\tau)}{r}=\evec_i\d_i\,\d_{i_1}\dots\d_{i_n}\frac{\Tsf_{i_1\dots i_n}(\tau)}{r},
\eeqan
 introduced for $r\,\ne\,0$ and with $\Tsft^{(n)}$ a {\bf STF} tensor . Let us consider the extension of this function as a distribution isolating the point-like singularities by the definition 
\beqa\label{65}
\left\langle\bbox{I}\left(\Tsft^{(n)}(t)\right),\,\phi\right\rangle&=&\lim_{\eps\to 0}\int_{\dom_\eps}\rmd^3x\,\left(\nablav^{n+1}\vert\vert\frac{\Tsft^{(n)}(\tau)}{r}\right)\,\phi(\rvec)\nonumber\\
&=&\evec_i\lim_{\eps\to 0}\int_{\dom_\eps}\rmd^3x\,\left(\d_i\,\d_{i_1}\dots \d_{i_n}\frac{\Tsf_{i_1\dots i_n}(\tau)}{r}\right)\,\phi(\rvec)\ .
\eeqa
The partial integration  becomes
\beqa\label{66}
\left\langle \bbox{I}\left(\Tsft^{(n)}(t)\right),\;\phi\right\rangle
 &=&\lim_{\eps\to 0}\left[\oint_{\Sigma_{\eps}}\rmd S\,\nuvec\left(\nablav^n\vert\vert\frac{\Tsft^{(n)}(\tau)}{r}\right)\phi(\rvec) \right. \nonumber\\
 &&\ \ \ \left. -\int_{\dom_\eps}\rmd^3x\,\left(\nablav^n\vert\vert\frac{\Tsft^{(n)}(\tau)}{r}\right)\,\nablav\phi(\rvec)
\right]\ .
\eeqa
 Considering only the surface integral from the last equation, employing equation \eref{5} and the Taylor series of the function $\phi(\rvec)$, after a regrouping of factors, we obtain
\beqa\label{67}
\fl\left\langle\,\bbox{I}_\sigma\left(\Tsft^{(n)}(t)\right),\;\phi\right\rangle&=&\evec_i\lim_{\eps\to 0}\oint_{\Sigma_\eps}\rmd\Omega(\nuvec)\suml^n_{l=0}\suml^\infty_{\al=0}\frac{\eps^{\al-l+1}}{c^{n-l}}\nonumber\\
\fl&&\ \ \ \times\nu_i\,C^{(n,l)}_{i_1\dots i_n}\frac{\rmd^{n-l}\Tsft_{i_1\dots i_n}(\tau_\eps)}{\rmd t^{n-l}}\nu_{i_{n+1}}\dots \nu_{i_{n+\al}}\left(\d_{i_{n+1}}\dots \d_{i_{n+\al}}\,\phi\right)_0 ,
\eeqa
where $\tau_\eps=r-\eps/c$. As in the static case, we can write this last expression in terms of the averages upon the directions:
\beqa\label{68}
\fl\left\langle\,\bbox{I}_\sigma\left(\Tsft^{(n)}(t)\right),\;\phi\right\rangle
&=&4\pi\,\evec_i\lim_{\eps\to 0}\suml^n_{l=0}\suml^\infty_{\al=0}\frac{\eps^{\al-l+1}}{c^{n-l}\al!}\frac{\rmd^{n-l}}{\rmd t^{n-l}}\left\langle\,C^{(n,l)}_{i_1\dots i_n}\Tsf_{i_1\dots i_n}(\tau_\eps)\nu_{i_{n+1}}\dots \nu_{i_{n+\al}}\,\nu_i\right\rangle\nonumber\\
\fl&&\ \ \ \times \left(\d_{i_{n+1}}\dots \d_{i_{n+\al}}\,\phi\right)_0 .
\eeqa
The general form of the coefficients $C^{(n,l)}$ is given by equation \eref{6} and one can notice that only the term $D^{(n,l)}_0\,\nu_{i_1}\dots \nu_{i_n}$  gives a non-zero contribution to the contraction with the {\bf STF} tensor $\Tsft^{(n)}$.  Therefore, we can write
\beqa\label{69}
\fl\left\langle\,\bbox{I}_\sigma\left(\Tsft^{(n)}(t)\right),\;\phi\right\rangle =
4\pi\,\evec_i\lim_{\eps\to 0}\suml^n_{l=0}\suml^\infty_{\al=0}\frac{\,\eps^{\al-l+1}}{c^{n-l}\al!}D^{(n,l)}_0\nonumber\\
 \times\frac{\rmd^{n-l}}{\rmd t^{n-l}}
\Tsf_{i_1\dots i_n}(\tau_\eps)\left\langle\,\nu_{i_1}\dots \nu_{i_n}\,\nu_{i_{n+1}}\dots \nu_{i_{n+\al}}\,\nu_i\right\rangle\,\left(\d_{i_{n+1}}\dots \d_{i_{n+\al}}\,\phi\right)_0
\eeqa
 \par In this equation, the factor $\eps^{\al-l+1}$ represents a negative power of $\eps$  for $\al\,<\,l-1$. Since $l\,\le \,n $, we can write
\beqa\label{70}
 \al\,<\,n-1\ .
 \eeqa
 In this case  
\beqa\label{71}
\left\langle\,\nu_{i_1}\dots \nu_{i_n}\,\nu_{i_{n+1}}\dots \nu_{i_{n+\al}}\,\nu_i\right\rangle\frac{\rmd^{n-l}}{\rmd t^{n-l}}\Tsf_{i_1\dots i_n}(\tau_\eps)=0 ,
\eeqa
 since $n+\al+1\,< \,2n$. Finally, the combinations of the factors $\eps^{\al-l+1}$ for $\al\,<\,n-1$ with the terms from the Taylor series of $\Tsf_{i_1\dots i_n}(\tau_\eps)$,
 \beqan
 \Tsf_{i_1\dots i_n}(\tau_\eps)=\suml^\infty_{\la=0}\frac{(-1)^\la\eps^\la}{\la!\,c^\la}\frac{\rmd^\la}{\rmd t^\la}\Tsf_{i_1\dots i_n}(t)
  \eeqan
  can give non negative powers of $\eps$, but equation \eref{71} remains valid.\\
  The terms not containing $\eps$ as factor verify the equality
\beqa\label{72}
\al\,=\,l-1
\eeqa
and they can give contribution different from  zero.
\par For $\al>\,l-1$ the corresponding terms contain positive powers of $\eps$ and have null  limits for $\eps\,\to\,0$. Consequently, the sum over $\al$ is limited at $\al=l-1$.
\par Summarising, one can write the following expression for the limit in equation \eref{66}:
\beqa\label{73}
\fl\left\langle\,\bbox{I}_\sigma\left(\Tsft^{(n)}(t)\right),\;\phi\right\rangle&=&4\pi\evec_i\suml^n_{l=1}\suml^{l-1}_{\al=0}\frac{D^{(n,l)}_{(0)}}{c^{n-l}\,\al!}\nonumber\\
\fl&\times&\frac{\rmd^{n-l}}{\rmd t^{n-l}}\Tsf_{i_1\dots i_n}(t)\left\langle\,\nu_{i_1}\dots\nu_{i_n}\nu_{i_{n+1}}\dots \nu_{i_{n+\al}}\,\nu_i\right\rangle\,\left(\d_{i{n+1}}\dots\d_{i_{n+\al}}\,\phi\right)_0 .
\eeqa
For $l\,<\,n$, equation \eref{72} becomes $\al\,<\,n-1$ and equation \eref{71} is verified; in the sum from equation \eref{73} remain only the terms corresponding to $l=n$ and $\al=n-1$ which give:
\beqa\label{72a}
\fl\left\langle\,\bbox{I}_\sigma\left(\Tsft^{(n)}(t)\right),\;\phi\right\rangle&=&\frac{(-1)^n4\pi (2n-1)!!}{(n-1)!}\nonumber\\
\fl&\times&\Tsf_{i_1\dots i_n}(t)\left\langle\,\nu_{i_1}\dots\nu_{i_n}\nu_{i_{n+1}}\dots\nu_{i_{2n-1}}\,\nu_i\right\rangle
\,\left(\d_{i_{n+1}}\dots \d_{i_{2n-1}}\,\phi\right)_0\ .
\eeqa
Comparing this result with the corresponding distribution \eref{37a} from the static case, one can see that equation \eref{72a} can be obtained from equation \eref{42a}  by the substitution $\Tsft^{(n)}\,\to\,\Tsft^{(n)}(t)$.
\par The limit of the volume integral from equation \eref{65} can be written  as
\beqa\label{73a}
\fl\left\langle\,J\left(\Tsft^{(n)}\right),\;\nablav\phi\right\rangle=\lim\int_{\dom_\eps}\rmd^3x\,\left(\nablav^n\frac{\Tsft^{(n)}(\tau)}{r}\right)\,\nablav\phi(\rvec)\\
\fl=\lim_{\eps\to 0}\left\{\oint_{\Sigma_\eps}\rmd S\left[\nuvec\cdot\left(\nablav^{n-1}\vert\vert\frac{\Tsft^{(n)}(\tau)}{r}\right)\right]\,\nablav\phi(\rvec)-\int_{\dom_\eps}\rmd^3x\,\left(\nablav^{n-1}\vert\vert\frac{\Tsft^{(n)}(\tau)}{r}\right)\vert\vert\nablav^2\phi(\rvec)\right\} .\nonumber
\eeqa
The limit of the surface integral from the last equation can be processed as in the case of equation \eref{67}, by writing
\beqa\label{74}
\fl &&\left\langle\,J_\sigma\left(\Tsft^{(n)}\right),\,\nablav\phi\right\rangle\nonumber\\
\fl&=&\evec_i\lim_{\eps\to 0}\oint_{\Sigma_\eps}\rmd S\,\nu_{i_n}\left(\d_{i_1}\dots \d_{i_{n-1}}\frac{\Tsf_{i_1\dots i_n}(\tau)}{r}\right)\,\d_i\phi(\rvec)
=\evec_i\lim_{\eps\to0}4\pi\,\suml^{n-1}_{l=0}\suml^\infty_{\al=0}\frac{(-1)^\la\,\eps^{\al-l+1}}{c^{n-l-1}\al!}\nonumber\\
\fl&\times&\left\langle\,C^{(n-1,n-1)}_{i_1\dots i_{n-1}}\nu_{i_n}\nu_{i_{n+1}}\dots\nu_{i_{n+\al}}\right\rangle
\frac{\rmd^{n-l-1}}{\rmd t^{n-l-1}}\Tsf_{i_1\dots i_n}(\tau_\eps)
\left(\d_i\d_{i_{n+1}}\dots \d_{i_{n+\al}}\,\phi\right)_0\ .
\eeqa
Only the terms with $\al-l+1=0$ can give contributions different from zero but, since $l\le n-1$, $\al$ verifies the inequality $\al\,\le\,n-2$. Consequently, the average from the last equation contains at most a symbol $\delta_{i_ji_k}$ with $j,\,k \le \,n$ and the contraction of this average with the {\bf STF} tensor $\Tsft^{(n)}$ is zero. Finally, by reccurence, one can verify that the limit \eref{73a} is zero such that equation 
\eref{72a} is the result for the limit from equation \eref{65}.
\par Similar considerations lead to the conclusion that the second term, proportional to $\nablav^{n-1}\vert\vert(\ddot{\psft}^{(n)}(\tau)/r )$,  in the right parenthesis from the expression \eref{59} of the electric field gives null contribution to the point-like singularities. The contribution of the magnetic moments to the point-like singularities of the electric field are defined by a distribution of the type 
\beqa\label{75}
\hspace{-0.5cm} \left\langle \bbox{L}(\Tsft^{(n)}),\;\phi\right\rangle &=& \evec_i\lim_{\eps\to 0}\,\eps_{ijk}\int_{\dom_\eps}\rmd^3x\,\left(\d_j\,\d_{i_1}\dots\d_{i_{n-1}}\frac{\Tsf_{i_1\dots i_{n-1}\,k}(\tau)}{r}\right)\,\phi(\rvec)\nonumber\\
&=&\evec_i\lim_{\eps\to 0}\,\left[\eps_{ijk}\oint_{\Sigma_\eps}\rmd S\,\nu_j\left(\d_{i_1}\dots\d_{i_{n-1}}\frac{\Tsf_{i_1\dots i_{n-1}\,k}(\tau)}{r}\right)\,\phi(\rvec)\right.\nonumber\\
&& \ \ \ -\eps_{ijk}\left.\int_{\dom_\eps}\,\rmd^3x\left(\d_{i_1}\dots\d_{i_{n-1}}\frac{\Tsf_{i_1\dots i_{n-1}\,k}(\tau)}{r}\right)\,\d_j\phi(\rvec)\right]\ .
\eeqa
Concerning the surface integral from equation \eref{75}, introducing, as in previous cases, equation \eref{5} for partial derivatives and the Taylor series for  $\phi(\rvec)$ (with the same notation for the various parameters), we notice that each term contains the contraction 
\beqan
\eps^{\al-l+1}\,\eps_{ii_nk}\left\langle\,C^{(n-1,n-1)}_{i_1\dots i_{n-1}}\,\nu_{i_n}\nu_{i_{n+1}}\dots \nu_{i_{n+\al}}\right\rangle\,\Tsf_{i_1\dots i_{n-1}\,k}(\tau_\eps)\ .
\eeqan
This contraction can give contributions different from zero only for $\al=l-1\,\le\,n-2$. 
As one can easily see, for such values of $\al$ this contraction vanishes either because of the presence of the symbols $\delta_{i_ni_q}$ with $q\le n-1$ or of $\delta_{i_qi_s}$ with $q,s\,\le\,n-1$. The same conclusion applies to the contributions of the electric moments to the magnetic field as seen from equation \eref{61}.
\par The last type of singular distribution we have to search is for the term $\Delta[\nablav^{n-1}\vert\vert(\Tsft^{(n)}(\tau)/r)]$ 
from the expansion \eref{61} of the magnetic field. The corresponding point-like singular distribution can be obtained from the definition
\beqa\label{76}
\hspace{-0.5cm} \left\langle\,\bbox{Q}\left(\Tsft^{(n)}\right),\;\phi\right\rangle &=& \evec_k\lim_{\eps\to 0}\int_{\dom_\eps}\rmd^3x\,\d_{i_1}\dots \d_{i_{n-1}}\,\left(\Delta\frac{\Tsf_{i_1\dots i_{n-1}\,k}(\tau)}{r}\right)\,\phi(\rvec)\nonumber\\
 &=& -4\pi\lim_{\eps\to 0}\int_{\dom_\eps}\rmd^3x\,\Tsf_{i_1\dots i_{n-1}\,k}(t)\,\phi(\rvec)\,\d_{i_1}\dots \d_{i_{n-1}}\,\delta(\rvec)\nonumber\\
 &=& -4\pi\,\Tsf_{i_1\dots i_{n-1}\,k}(t)\,\left\langle\,\d_{i_1}\dots \d_{i_{n-1}}\,\delta,\;\phi\right\rangle\ .
\eeqa
Equations \eref{72a} and \eref{76}, inserted in equation \eref{61}, lead to the same conclusion as in the electric case: the distribution $\Bvec^{(n)}_{(0)}$ in the dynamic case can be obtained from the static case, given by equation \eref{53a}, with the substitution $\mct^{(n)}\,\to\,\mct^{(n)}(t)$. 

\section{Conclusion}
Based on the results of the previous section, it is obvious that in the expansions \eref{59} and \eref{61} of $\Evec(\rvec,t)$ and $\Bvec(\rvec,t)$, after performing the substitutions \eref{62}, only the terms
\beqan
\nablav\left(\nablav^{\,n}\vert\vert\frac{\psftr^{(n)}(\tau)}{r}\right)
\eeqan
for the electric field, and
\beqan
\nablav\left(\nablav^{\,n}\vert\vert\frac{\msftr^{(n)}(\tau)}{r}\right),\;\;\;
\Delta\left(\nablav^{n-1}\vert\vert\frac{\msftr^{(n)}(\tau)}{r}\right)
\eeqan
for the magnetic one give contributions to the $\delta$-singularities of the electromagnetic field. These contributions can be obtained by simply performing in the corresponding static expressions \eref{49} and \eref{53a} of $\Evec$ and $\Bvec$, the substitutions
\beqa\label{77}
\pct^{(n)}\,\to\,\psftr^{(n)}(t),\;\;\;\mct^{(n)}\,\to\,\msftr^{(n)}(t)\ .
\eeqa
In \ref{ap2}, the reduced {STF} moments for the first multipoles are established. Considering the substitutions \eref{77} performed for all the multipoles, obviously $\psftr^{(n)}(t)$ and $\msftr^{(n)}(t)$ are actually represented by infinite series. Practically, as seen from this appendix, for a given multipole approximation, one deals only with finite sums. If, for example, we are interested only in the contributions to the singularities of the field from the electric octupole ($n=3$) and magnetic quadrupole ($n=2$), besides the contributions of the {\bf STF} projections $\pct^{(3)}$ and $\mct^{(2)}$, contributions given by equations \eref{50} and \eref{56}, we have to consider also a contribution of the type \eref{31} of the electric dipolar toroidal  moment $\bbox{T}$ defined by equation \eref{A44}.  
This last contribution is given by equation \eref{31} with $\pvec\,\to\,\widetilde{\pvec}=\pvec-\dot{\bbox{T}}/c^2$. If we are interested only in the contribution of the magnetic quadrupole, besides the contribution of the corresponding {\bf STF} projection, we have to consider the contribution of an electric dipolar moment $\Delta'\pvec$ given by equation \eref{A40}, and so on.

\par We point out that there are alternative ways for introducing the moments $\psftr$ and $\msftr$. Obviously, the results from the present paper obtained through the formalism of the Cartesian tensors, can be obtained employing the expansions of the electromagnetic field in the formalism of the spherical tensors as in Refs.\  \cite{Jackson} and especially \cite{Dubovik-FEC, Dubovik-rep}.  One can employ the multipole moments derived by  this formalism  and then introduce the connection between spherical and Cartesian components of the multipoles \cite{RV} . 
\par The results of the present paper, if correct, could be useful in studying some problems of interest from the atomic, nuclear, elementary particles, and even gravitation physics.

 \appendix
 \section{Expressions of the coefficients $C^{(n,l)}$}
The coefficients $C$ from equation \eref{5} for $0\,\le\,n\,\le 4$ are:
\beqan
C^{0,0}=1\ ;
\eeqan
\beqa\label{C001}
C^{(1,0}_i=-\nu_i,\;\;\;C^{1,1)}_i=-\nu_i\ ;
\eeqa
\beqa\label{C002}
C^{(2,0)}_{ij}=\nu_i\nu_j,\;\;C^{2,1}_{ij}=3\nu_i\nu_j-\delta_{ij},\;\;\;C^{(2,2)}_{ij}=3\nu_i\nu_j-\delta_{ij}\ ;
\eeqa
\beqa\label{C003}
&~&C^{(3,0)}_{ijk}=-\nu_i\nu_j\nu_k,\;\;\;C^{(3,1}_{ijk}=-6\nu_i\nu_j\nu_k+\delta_{\{ij}\nu_{k\}},\nonumber\\
&~&C^{(3,2)}=-\,15\nu_i\nu_j\nu_k+3\,\delta_{\{ij}\nu_{k\}},\;\;C^{(3,3)}=-\,15\nu_i\nu_j\nu_k+3\,\delta_{\{ij}\nu_{k\}};
\eeqa
\beqa\label{C004}
 C^{(4,0)}_{ijkl}=\nu_i\nu_j\nu_k\nu_l,\nonumber\\
 C^{(4,1)}_{ijkl}=10\nu_i\nu_j\nu_k\nu_l-\,\delta_{\{ij}\nu_k\nu_{l\}},\nonumber\\
 C^{(4,2)}_{ijkl}=45\,\nu_i\nu_j\nu_k\nu_l-\,6\delta_{\{ij}\nu_k\nu_{l\}}+\,\delta_{\{ij}\delta_{kl\}},\nonumber\\
 C^{(4,3)}_{ijkl}=105\,\nu_i\nu_j\nu_k\nu_l-15\,\delta_{\{ij}\nu_k\nu_{l\}}+3\,\delta_{\{ij}\delta_{kl\}},\nonumber\\
 C^{(4,4)}=105\,\nu_i\nu_j\nu_k\nu_l-15\,\delta_{\{ij}\nu_k\nu_{l\}}+3\,\delta_{\{ij}\delta_{kl\}}.
\eeqa

\section {Reduction of the multipole tensors}
\label{ap2}
\par Let us consider the electric quadripolar and magnetic dipolar contributions to the electromagnetic fields. The 4-polar electric moment $\psft^{(2)}$ represents the first tensor for which one has to apply the substitution  by the corresponding {\bf STF} one. The tensor $\psft^{(2)}$ is symmetric and the corresponding trace-free projection, denoted by $\pct^{(2)}$  can be  established as a combination of the form
\beqa\label{A28}
\pc_{ij}=\psf_{ij}-\Lambda\delta_{ij}.
\eeqa
The parameter $\Lambda$ is determined from the condition $\pc_{jj}=0$:
\beqa\label{La}
\Lambda=\frac{1}{3}\psf_{ii}\ .
\eeqa
 Firstly, we observe that $\Evec$ and $\Bvec$ are not changed by the substitution $\psft^{(2)}\,\to\,\pct^{(2)}$. Indeed,  equation \eref{55} reads:
\beqa\label{A29}
\Evec(\rvec,t)\;\stackrel{\psft^{(2)}\,\to\,\pct^{(2)}}{\,\longrightarrow}\;\Evec(\rvec,t) +\frac{1}{8\pi\eps_0}\evec_i\left[\Delta\d_i\frac{\Lambda(\tau)}{r}-\frac{1}{c^2}\d_i\frac{\ddot{\Lambda}(\tau)}{r}\right]\ .
\eeqa
Since $r\ne 0$,  equation \eref{A29} shows the invariance of the electric field to the considered substitution . Considering the change in the expression \eref{57} by the substitution $\psft^{(2)}\,\to\,\pct^{(2)}$, we obtain 
\beqa\label{A29a}
\hspace{-1.5cm} \Bvec(\rvec,t)\,\stackrel{\psft^{(2)}\,\to\,\pct^{(2)}}{\,\longrightarrow}\Bvec(\rvec,t)+\frac{\mu_0}{8\pi}\evec_i\,\eps_{ijk}\d_j\d_l\delta_{lk}\frac{\dot{\Lambda}(\tau)}{r}=\frac{\mu_0}{8\pi}\evec_i\eps_{ijk}\,\d_j\d_k\frac{\dot{\Lambda}(\tau)}{r}=0\ ,
\eeqa
$\Bvec$ being also invariant to this substitution.
Let us consider the substitution 
\beqa\label{A32}
\psft^{(3)}\,\to\,\pct^{(3)}
\eeqa
in the multipole expansions of the fields $\Evec$ and $\Bvec$. We have to express the components $\pc_{ijk}$ in terms of the components $\psf_{ijk}$. Obviously, the relations between the components of the tensors $\psft^{(3)}$ and $\pct^{(3)}$ are of the following form:
\beqa\label{A33}
\pc_{ijk}=\psf_{ijk}-\delta_{\{ij}\Lambda_{k\}}.
\eeqa
Considering the conditions on  the traces $\pc_{iij}$ we write
\beqan
\pc_{iij}=\psf_{iij}-5\,\Lambda_j=0,
\eeqan
i.e.\  
\beqa\label{lai}
\Lambda_i=\frac{1}{5}\psf_{ijj}=\frac{1}{5}\int_{\dom}\rmd^3x\,r^2\,x_i\,\rho \ .
\eeqa
The effect of the substitution \eref{A32} in the expansion \eref{55} of $\Evec$ is given by
\beqa\label{A34}
\fl\Evec^{(3)}_p(\rvec,t)\,&\to &\Evec^{(3)}_p(\rvec,t)-\frac{1}{24\pi\eps_0}\,\evec_i\left[\d_i\d_j\d_k\d_l\frac{\delta_{\{kl}\Lambda_{j\}}(\tau)}{r}-\frac{1}{c^2}\d_k\d_l\frac{\delta_{\{kl}\ddot{\Lambda}_{i\}}}{r}\nonumber\right]\\
\fl&=&\Evec^{(3)}_p(\rvec,t)-\frac{1}{24\pi\eps_0c^2}\left[\evec_i\,\d_i\d_j\frac{\ddot{\Lambda}_j(\tau)}{r}+\frac{1}{c^2}\evec_i\frac{\qdot{\Lambda}_i(\tau)}{r}\right],
\eeqa
or with the tensorial notation, the change of the field $\Evec$ is given by
\beqa\label{A35}
\Evec(\rvec,t)\,\stackrel{\psft^{(3)}\to \pct^{(3)}}{\longrightarrow}\;\Evec(\rvec,t)-\frac{1}{24\pi\eps_0c^2}\left[\nablav\left(\nablav\cdot\frac{\Laop(\tau)}{r}\right)-\frac{1}{ c^2}\frac{\qdot{\Laop}(\tau)}{r}\right],
\eeqa
where $\Laop=\evec_i\,\Lambda_i$. This expression of the modification of $\Evec$ suggests us to try a change of the electric dipolar moment $\pvec$ by which the modification produced by the substitution \eref{A32} can be compensated. 
Let be the substitution:
\beqa\label{A35a}
\pvec\,\to\, \pvec'=\pvec+\Delta \pvec .
\eeqa
We try to determine $\Delta\pvec$ such that the modification \eref{35} of $\Evec$ is compensated. Since
\beqa\label{A35b}
\Evec(\rvec,t)&\stackrel{\pvec\to\Delta\pvec}{\longrightarrow}&\,\Evec(\rvec,t)+\frac{1}{4\pi\eps_0}\evec_i\left[\d_i\d_j\frac{\Delta p_j(\tau)}{r}-\frac{1}{c^2}\frac{\Delta \ddot{p}_j(\tau)}{r}\right]\nonumber\\
&=&\Evec(\rvec,t)+\frac{1}{4\pi\eps_0c^2}\left[\nablav\left(\nablav\cdot\frac{\Delta\pvec(\tau)}{r}\right)-\frac{1}{c^2}\frac{\Delta\ddot{\pvec}}{r}\right],                          
\eeqa
the comparison with equation \eref{A35} gives  $\Delta\pvec=\ddot{\Laop}/6c^2$ and the substitution is in fact
\beqa\label{A36}
\pvec\,\to\,\pvec'=\pvec+\frac{1}{6c^2}\,\ddot{\Laop}\ .
\eeqa
Let us consider also the change of $\Bvec$ by the substitutions \eref{A32} and \eref{A36}:
\beqan
\fl&&\Bvec(\rvec,t)\stackrel{\pvec\to\pvec',\,\psft^{(3)}\to\pct^{(3)}}{\longrightarrow}\,\Bvec(\rvec,t)+\frac{\mu_0}{4\pi}\evec_i\eps_{ijk}\,\d_j\frac{\Delta\dot{p}_k(\tau)}{r}-\frac{\mu_0}{24\pi}\evec_i\eps_{ijk}\,\d_j\d_l\d_q\,
\frac{\delta_{\{lq}\dot{\Lambda}_{k\}}(\tau)}{r}\nonumber\\
\fl&=&\Bvec(\rvec,t)+\frac{\mu_0}{24\pi c^2}\evec_i\eps_{ijk}\d_j\frac{\tdot{\Lambda}_k(\tau)}{r}-\frac{\mu_0}{24\pi}\evec_i\eps_{ijk}\,\d_j\d_l\d_q\frac{\left(\delta_{lq}\dot{\Lambda}_k(\tau)+\delta_{lk}\dot{\Lambda}_q(\tau)+\delta_{qk}\dot{\Lambda}_l(\tau)\right)}{r}\nonumber\\
\fl&=&\Bvec(\rvec,t)+\frac{\mu_0}{24\pi c^2}\evec_i\eps_{ijk}\d_j\frac{\tdot{\Lambda}_k(\tau)}{r}-\frac{\mu_0}{24\pi}\evec_i\eps_{ijk}\d_j\left(\Delta\frac{\dot{\Lambda}_k(\tau)}{r}+2\,\d_k\d_l\frac{\dot{\Lambda}_l(\tau)}{r}\right)=0.
\eeqan 
Since
\beqan
\Delta\frac{\dot{\Lambda}_k(\tau)}{r}=\frac{\tdot{\Lambda}_k(\tau)}{c^2\,r}\;\;\mbox{and}\;\;\eps_{ijk}\,\d_j\d_k=0,
\eeqan
 $\Bvec$ is invariant to such a substitution. Concluding, we can substitute the  tensor $\pct^{(3)}$ by the {\bf STF} tensor if we perform simultaneously the substitution $\pvec\,\to\,\pvec'=\pvec+\ddot{\Lambda}/{6c^2}$.
\par Since we have to consider, together with the contributions of the electric octupolar moment to the $\delta$-singularities of the field, such contributions of the magnetic quadripolar moment also, we must show how one can substitute the momentum tensor $\msft^{(2)}$ by the corresponding {\bf STF} projection $\mct^{(2)}$. Let be the components 
\beqan
\msf_{ij}(t)=\frac{2}{3}\int_{\dom}\rmd^3x\,x_i\left(\rvec\times\jvec(\rvec,t)\right)\ .
\eeqan
Concerning the magnetic quadrupole moment $\msf^{(2)}$, we have a simple procedure for obtaining the {\bf STF} projection (up to a factor). Let us write the identity:
\beqan
\msf_{ij}=\frac{1}{2}(\msf_{ij}+\msf_{ji})+\frac{1}{2}(\msf_{ij}-\msf_{ji}),
\eeqan
where the first bracket represents the symmetric part of this tensor, and the second, the antisymmetric one. The symmetric part is, for this case ($n=2$), a {\bf STF} tensor 
$\mct^{(2)}$. Therefore, 
\beqa\label{A37}
\msf_{ij}=\mc_{ij}+\frac{1}{2}\,\eps_{ijk}\,\Nsf_k,
\eeqa 
where 
   \beqa\label{A38}
\fl\;\;\;\;\;\;\;\;\;\;\Nsf_k&=&\eps_{kij}\,\msf_{ij}=\frac{2}{3\al}\int_{\dom}\big[\rvec\times(\rvec\times\jvec)\big]_k\,\rmd^3x
=\frac{2}{3\al}\int_{\dom}\big[(\rvec\cdot\jvec)\,x_k-r^2\,J_k\big]\,\rmd^3x.
\eeqa
The substitution $\msft^{(2)}\,\to\,\mct^{(2)}$ in equation \eref{55} gives:
\beqa\label{A39}
\Evec(\rvec,t)\,&&\stackrel{\msft^{(2)}\,\to\,\mct^{(2)}}{\longrightarrow}\Evec(\rvec,t)-\frac{1}{16\pi\eps_0c^2}\evec_i\,\eps_{ijk}\eps_{lkq}\,\d_j\d_l\frac{\dot{\Nsf}_q(\tau)}{r}\nonumber\\
&=&\Evec(\rvec,t)+\frac{1}{16\pi\eps_0c^2}\left[\nablav\left(\nablav\cdot\frac{\dot{\bbox{N}}(\tau)}{r}\right)-\Delta\frac{\dot{\bbox{N}}(\tau)}{r}\right]\nonumber\\
&=&\Evec(\rvec,t)+\frac{1}{16\pi\eps_0c^2}\left[\nablav\left(\nablav\cdot\frac{\dot{\bbox{N}}(\tau)}{r}\right)-\frac{\tdot{\bbox{N}}(\tau)}{c^2r}\right],
\eeqa
where $\bbox{N}=\Nsf_i\evec_i$. If we perform the substitution of the electric dipolar moment $\pvec\,\to\,\pvec''=\pvec+\Delta'\pvec$, the electric field $\Evec$ is modified by an expression given by equation \eref{A35b} (with $\Delta\to\Delta'$). It is easy to see that  the compensation of the changes in the electric field produced by the substitutions $\msft^{(2)}\,\to\,\mct^{(2)}$ and $\pvec\,\to\,\pvec''=\pvec+\Delta'\pvec$, is given by the choice:
\beqa\label{A40}
\Delta'\pvec=-\frac{1}{4c^2}\dot{\bbox{N}}\ .
\eeqa
Concerning the change of $\Bvec$ by the substitutions $\pvec\,\to\,\pvec+\Delta'\pvec,\;\msft^{(2)}\,\to\,\mct^{(2)}$, 
we can write:
\beqa\label{A41}
\fl\Bvec(\rvec,t) && \stackrel{\pvec\to\pvec+\Delta'\pvec,\;\msft^{(2)}\to\mct^{(2)}}{\longrightarrow}\Bvec(\rvec,t)+\frac{\mu_0}{4\pi}\nablav\times\frac{\Delta'\dot{\pvec}(\tau)}{r}\nonumber\\
\fl&& \hspace{5cm} +\frac{\mu_0}{16\pi}\evec_i\left[\eps_{kjl}\d_i\d_j\d_k\,\frac{\Nsf_l(\tau)}{r}-\eps_{kil}\d_k\Delta\frac{\Nsf_l(\tau)}{r}\right]\nonumber\\
\fl&=&\Bvec(\rvec,t)
-\frac{1}{16\pi c^2}\,\nablav\times\frac{\ddot{\bbox{N}}(\tau)}{r}-\frac{\mu_0}{16\pi}\left[\nablav\left(\nablav\cdot\big(\nablav\times\frac{\bbox{N}(\tau)}{r}\big )\right)-\frac{1}{c^2}\nablav\times\frac{\ddot{\bbox{N}}(\tau)}{r}\right]\nonumber\\
\fl&=&\Bvec(\rvec,t),
\eeqa
i.e.\  the invariance of $\Bvec$.
Finally, we can express the total effect of the two substitutions $\psft^{(3)}\,\to\,\pct^{(3)}$ and $\msft^{(2)}\,\to\,\mct^{(2)}$ represented by the transformation of the electric dipolar moment:
\beqa\label{A42}
\pvec\,\stackrel{\psft^{(3)}\,\to\,\pct^{(3)}}{\longrightarrow}\,\pvec+\frac{1}{6c^2}\,\ddot{\Laop}\,
\stackrel{\msft^{(2)}\,\to\,\mct^{(2)}}{\longrightarrow}\,\widetilde{\pvec}=\pvec+\frac{1}{6c^2}\ddot{\Laop}-\frac{1}{4c^2}\dot{\bbox{N}},
\eeqa
or
\beqa\label{A43}
\pvec\,\stackrel{\psft^{(3)}\,\to\,\pct^{(3)},\;}{\longrightarrow}\stackrel{\msft^{(3)}\,\to\,\mct^{(3)}}{\longrightarrow}\,\widetilde{\pvec}=\pvec-\frac{1}{c^2}\dot{\bbox{T}},
\eeqa
where 
\beqa\label{A44}
\bbox{T}=\frac{1}{4}\bbox{N}-\frac{1}{6}\dot{\Laop}
\eeqa
is the electric dipolar toroidal moment \cite{Dubovik-FEC}.

\end{document}